\documentclass{elsarticle}\usepackage[]{graphicx}\usepackage[]{color}
\pdfoutput=1
\makeatletter
\def\maxwidth{ %
  \ifdim\Gin@nat@width>\linewidth
    \linewidth
  \else
    \Gin@nat@width
  \fi
}
\makeatother

\definecolor{fgcolor}{rgb}{0.345, 0.345, 0.345}

\usepackage{framed}
\makeatletter
 {\par\unskip\endMakeFramed%
 \at@end@of@kframe}
\makeatother

\definecolor{shadecolor}{rgb}{.97, .97, .97}
\definecolor{messagecolor}{rgb}{0, 0, 0}
\definecolor{warningcolor}{rgb}{1, 0, 1}
\definecolor{errorcolor}{rgb}{1, 0, 0}
\newenvironment{knitrout}{}{} % an empty environment to be redefined in TeX

\usepackage{alltt}

\usepackage[utf8]{inputenc} % pacote para acentuação

\PassOptionsToPackage{hyphens}{url}
\usepackage{hyperref}
\expandafter\def\expandafter\UrlBreaks\expandafter{\UrlBreaks%  save the current 一
  \do\a\do\b\do\c\do\d\do\e\do\f\do\g\do\h\do\i\do\j%
  \do\k\do\l\do\m\do\n\do\o\do\p\do\q\do\r\do\s\do\t%
  \do\u\do\v\do\w\do\x\do\y\do\z\do\A\do\B\do\C\do\D%
  \do\E\do\F\do\G\do\H\do\I\do\J\do\K\do\L\do\M\do\N%
  \do\O\do\P\do\Q\do\R\do\S\do\T\do\U\do\V\do\W\do\X%
  \do\Y\do\Z\do\1\do\2\do\3\do\4\do\5\do\6\do\7\do\8\do\9\do\0}

\usepackage{mathtools}          % for \DeclarePairedDelimiter
\usepackage{amssymb}
\DeclarePairedDelimiter{\floor}{\lfloor}{\rfloor}
\DeclarePairedDelimiter{\ceil}{\lceil}{\rceil}
\usepackage{graphicx}           % pacote para importar figuras
\usepackage{tikz}
\usetikzlibrary{decorations.pathreplacing}
\usetikzlibrary{shapes.misc}
\tikzset{cross/.style={cross out, draw,
         minimum size=2*(#1-\pgflinewidth),
         inner sep=0pt, outer sep=0pt}}

\newcommand{\drawolddomitem}[2]{
	\edef \x     {#1}
	\edef \y     {#2}
	\draw (\x, \y) node[cross=2.5pt,draw=gray] {};
}

\newcommand{\drawhvectorfill}[5]{
	\edef \origin {#1}
	\edef \xmax {#2}
	\edef \rs {(#3,#4)}
	\edef \filling {#5}
	\foreach \x in {0,...,\xmax}{
		\draw [shift={\origin},fill=\filling] (\x,0) rectangle +\rs;
	}
}
\newcommand{\drawrectangle}[4]{
	\edef \x     {#1}
	\edef \y     {#2}
	\edef \size  {#3}
	\edef \color {#4}
	\draw [fill=\color] (\x - \size, \y - \size) rectangle +(2*\size, 2*\size);
}
\newcommand{\drawaxis}[4]{
	\edef \xmax {#1}
	\edef \ymax {#2}
	\edef \xlab {#3}
	\edef \ylab {#4}
	\draw [<->, thick] (\xmax,0) -- (0,0) -- (0,\ymax);
	\node [below, font=\small] at (\xmax/2, 0) {\xlab};
	\node [rotate=90, above, font=\small] at (0,\xmax/2) {\ylab};
}

\usepackage{algorithm}
\usepackage{algpseudocode}

\usepackage{longtable}
\usepackage{pdflscape} % used for CSP experiment table
\usepackage{tabu} % the latex wikibook suggest this table over others
\usepackage{array} % for common commands inside \newcolumntype
\usepackage{makecell} % for makecell, that allows libreaks in cells

\usepackage{verbatim} % for \begin{comment} remove before final version
\usepackage{adjustbox}
\usepackage[nice]{nicefrac}
\usepackage{float}
\usepackage{tabularx}
\usepackage[section]{placeins}
\makeatletter
\AtBeginDocument{%
  \expandafter\renewcommand\expandafter\subsection\expandafter{%
    \expandafter\@fb@secFB\subsection
  }%
}
\makeatother

\title{An empirical analysis of exact algorithms for the\\unbounded knapsack problem}
\IfFileExists{upquote.sty}{\usepackage{upquote}}{}
\begin{document}

% Inside document definitions for tabu
\newcolumntype{L}[1]{>{\hsize=#1\hsize\raggedright\arraybackslash}X}
\newcolumntype{R}[1]{>{\hsize=#1\hsize\raggedleft\arraybackslash}X}
\newcolumntype{C}[1]{>{\hsize=#1\hsize\centering\arraybackslash}X}

\begin{frontmatter}
\author[ufrgs]{Henrique~Becker}
\ead{hbecker@inf.ufrgs.br}
\author[ufrgs]{Luciana~S.~Buriol}
\ead{buriol@inf.ufrgs.br}

%\cortext[cor1]{Principal corresponding author}
%\cortext[cor2]{Corresponding author}

\address[ufrgs]{Federal University of Rio Grande do Sul (UFRGS), Porto Alegre, Brazil}

\begin{abstract}
This work presents an empirical analysis of exact algorithms for the unbounded knapsack problem, which includes seven algorithms from the literature, two commercial solvers, and more than ten thousand instances.
The terminating step-off, a dynamic programming algorithm from 1966, presented the lowest mean time to solve the most recent benchmark from the literature.
The threshold and collective dominances are properties of the unbounded knapsack problem first discussed in 1998, and exploited by the current state-of-the-art algorithms. 
The terminating step-off algorithm did not exploit such dominances, but has an alternative mechanism for dealing with dominances which does not explicitly exploits collective and threshold dominances.
Also, the pricing subproblems found when solving hard cutting stock problems with column generation can cause branch-and-bound algorithms to display worst-case times.
The authors present a new class of instances which favors the branch-and-bound approach over the dynamic programming approach but do not have high amounts of simple, multiple and collective dominated items.
This behaviour illustrates how the definition of hard instances changes among algorithm approachs.
The codes used for solving the unbounded knapsack problem and for instance generation are all available online.
\end{abstract}

% NOTE: DO NOT ADD MORE THAN FIVE KEYWORDS, THE FIRST KEYWORD NEEDS TO BE
% FROM A SPECIAL LIST CONTAINED IN THE AUTHORS MATERIAL

\end{frontmatter}

%\begin{listofabbrv}{PRNG}
%        \item[B\&B] Branch and Bound
%        \item[BKP] Bounded Knapsack Problem
%        \item[BPP] Bin Packing Problem
%	\item[CA] Consistency Approach
%	\item[CPU] Central Processing Unit
%        \item[CSP] Cutting Stock Problem
%        \item[DP] Dynamic Programming
%	\item[FP] Floating Point
%	\item[GUKP] General Unconstrained Knapsack Problem
%	\item[GCKP] General Constrained Knapsack Problem
%        \item[KP] Knapsack Problem
%        \item[PRNG] Pseudo-Random Number Generator
%	\item[SCF] Set Covering Formulation
%	\item[SD] Standard Deviation
%        \item[UKP] Unbounded Knapsack Problem
%\end{listofabbrv}

% idem para a lista de símbolos
%\begin{listofsymbols}{$\alpha\beta\pi\omega$}
%       \item[$\sum{\frac{a}{b}}$] Somatório do produtório
%       \item[$\alpha\beta\pi\omega$] Fator de inconstância do resultado
%\end{listofsymbols}

% lista de figuras
%\listoffigures

% lista de tabelas
%\listoftables

% sumario
%\tableofcontents

% to make latex see the pdfs created by Sweave
%\graphicspath{{sections/}}

\textbf{The formal publication of this article can be found in \url{https://doi.org/10.1016/j.ejor.2019.02.011}.}

\copyright 2019. This manuscript version is made available under the CC-BY-NC-ND 4.0 license \url{http://creativecommons.org/licenses/by-nc-nd/4.0/}

\section{Introduction}
\label{sec:introduction}

The objective of this work is to provide an extensive comparison of the exact algorithms for solving the Unbounded Knapsack Problem (UKP).
Given the weight capacity of a knapsack and a collection of items (each with a weight and a profit value), the UKP consists in choosing how many copies of each item will be packed in the knapsack to maximize the profit carried by it while respecting its weight capacity.
The UKP is similar to the Bounded Knapsack Problem (BKP) and the 0-1 Knapsack Problem (0-1 KP).
The only difference between the UKP and the BKP (or the 0-1 KP) is that the UKP has an unlimited quantity of each item available.
The UKP is a weakly NP-Hard problem, as are the BKP and the 0-1 KP.
%~\cite{where_hard}.
%The UKP can also be seen as a special case of the BKP in which, for each item type, there are more copies available than is possible to fit in the knapsack capacity.

%\subsection{Prior work}

Next the authors present some selected papers and a summary of their relevance for the UKP.

\begin{table}[!htb]
\renewcommand{\arraystretch}{1.8}
\label{tab:prior_work}
%\begin{adjustbox}{max width=\textwidth, center}
\begin{tabu}{p{\textwidth}}
Summary of the prior work\\
\hline
\cite{gg-61} The column generation approach for the CSP linear programming relaxation is proposed; the UKP is the pricing problem.\\
\cite{gg-66} The ordered and the terminating step-off algorithms (dynamic programming algorithms to solve the UKP) are proposed. \\
\cite{mtu1} The MTU1 (branch-and-bound algorithm) is proposed, and then compared with the previous algorithms over artificial instances up to a hundred items, obtaining slightly better results \\
\cite{mtu2} Datasets of instances with up to 250,000 items (but rich in simple and multiple dominances) are proposed. MTU2 is proposed as an improvement of MTU1 for such dataset.\\
%1997 & \cite{zhu_dominated} & Uncorrelated datasets are shown to  \\
%1997 & \cite{babayev} & A new solving approach, more similar to dynamic programming than B\&B \\
\cite{ukp_new_results, eduk} EDUK (a dynamic programming algorithm) is proposed. Threshold dominance is proposed. EDUK is the first algorithm to exploit collective and threshold dominances. Old datasets receive some criticism for their high percentage of dominated items. New artificial datasets without the same flaws of the previous datasets are proposed. EDUK is compared to MTU2.\\
%2000 & \cite{eduk} & The new DP method only compares to B\&B and naive DP, the old non-naive DP algorithms were forgotten or excluded because of previous experiments. \\
\cite{book_ukp_2004} A book on knapsack problems cite EDUK as state-of-the-art dynamic programming for the UKP. \\
\cite{pya} EDUK2 is proposed. It consists in EDUK hybridized with B\&B. The datasets are updated to be `harder'. MTU2 is used in some comparisons, but not in all because it has the risk of integer overflow. EDUK2 is compared to EDUK.\\ %Such datasets are hard for B\&B, and the hybrid method is only compared to B\&B. The hybrid method is the new state-of-the-art. \\
\cite{sea2016} The terminating step-off is reinvented (with the name of UKP5) and outperforms the hybrid method in the updated datasets. \vspace{2mm}\\
\hline
\end{tabu}
%\end{adjustbox}
\end{table}

The UKP is the pricing subproblem generated by solving the linear programming relaxation of the set covering formulation for the unidimensional Bin Packing Problem (BPP) and Cutting Stock Problem (CSP) using the column generation approach~\citep{gg-61,gg-63}.
The BPP and the CSP are classical optimization problems in the area of operations research~\citep{survey2014}.
The best lower bounds known for the BPP and CSP are their linear programming relaxations of the set covering formulation.
This is the tightest formulation for these problems but it has an exponential number of columns, and thus is solved by using the column generation approach~\citep{gg-61}.
However, recently \cite{eq_lb_delorme} proved that a pseudo-polynomial formulation (dynamic programming-flow formulation) is equivalent to the set covering formulation and, therefore, provides the same lower bounds for the problem.

The main contributions made by this paper follow:

\begin{enumerate}
\item a comprehensive empiric analysis including several algorithms, solvers, and datasets, in a single self-contained paper -- in \autoref{sec:exp_and_res};
\item an updated pseudocode of the ordered step-off~\citep{gg-66}, using the current nomenclature and a new tiebreaker -- in \autoref{sec:oso_and_sol_dom};
\item the concept of \emph{solution dominance} which generalizes all previously proposed dominances -- in \autoref{sec:oso_and_sol_dom};
\item the evidence that \emph{partial solution dominance} is a competitive alternative to the state-of-the-art application of simple, multiple, collective, and threshold dominances -- in Section \autoref{sec:exp_and_res};
\item the discussion on how tighter bounds for periodicity do not help to improve the performance of state-of-the-art algorithms -- in \autoref{sec:critique_of_periodicity};
\item a new item distribution and how it favors a solving approach in relation to other (B\&B over DP) -- in sections \ref{sec:breq_inst} and \ref{sec:breq_exp};
\item a study of the item distribution evolution in CSP/BPP pricing problems and some of its implications (how it biases against some UKP-solving approaches, how it is affected by `hard' CSP/BPP instances) -- in \autoref{sec:csp_experiments};
\end{enumerate}

In the remainder of this section we discuss the mathematical formulation and well-known properties of the problem.

\subsection{Formulation and notation}
\label{sec:formulation}

An instance of the UKP is a pair of a capacity~\(c\) and a list of~\(n\) items.
Each item~\(i\) can be referenced by its index in the item list~\(i \in \{1,~\dots,~n\}\), and has a weight value~\(w_i\), and a profit value~\(p_i\).
A solution is an item multiset (i.e., a set that allows multiple copies of the same element).
The sum of the items weight, or profit, of a solution~\(s\) is denoted by~\(w_s\), or~\(p_s\), and is referred to as the weight, or profit, of the solution.
%A solution~\(s\) is valid iff~\(w_s \leq c\).
%An optimal solution~\(s^*\) is a valid solution with the greatest profit value among all valid solutions.
%To solve an instance of the UKP is to find an optimal solution for that instance.
%The profit value shared by all optimal solutions is denoted by \(p_{s^*}\).
%We refer to the profit value shared among all optimal solutions for a capacity~\(y\) as~\(opt(y)\), if we omit the capacity then~\(c\) is implied.

The mathematical formulation of UKP is:

\begin{align}
  \mbox{maximize} &\sum_{i=1}^n p_i x_i\label{eq:objfun}\\
\mbox{subject~to} &\sum_{i=1}^n w_i x_i \leq c\label{eq:capcons},\\
            &x_i \in \mathbb{N}_0.\label{eq:x_integer}
\end{align}

The quantities of each item~\(i\) in a solution are denoted by~\(x_i\), and are restricted to the non-negative integers, as~Eq.~\eqref{eq:x_integer} indicates. 
We assume that the capacity~\(c\), the quantity of items~\(n\) and the weights of the items~\(w_i\) are positive integers, while the profit values of the items~\(p_i\) are positive real numbers.

The efficiency of an item~\(i\) is its profit-to-weight ratio (\(p_i/w_i\)).%, and is denoted by~\(e_i\). 
We use~\(w_{min}\), or~\(w_{max}\), to denote the lowest weight among all items, or the highest weight among all items, within an instance of the UKP.
We refer to the item with the greatest efficiency among all items of a specific instance as the \emph{best item} (or~\(b\)). If more than one item shares the greatest efficiency, then the item with the lowest weight among them is considered the best item type. 
If more than an item has both previously stated characteristics (i.e., they are equal), then the first item with both characteristics in the items list is the best item.
The authors make such distinction because the instance generators do not guarantee unique items, and is often faster to let the algorithms themselves deal with the replicas than running a preprocessing phase to remove duplicated items.

\subsection{Properties of the UKP}
\label{sec:well_known_prop}

Algorithms that solve the UKP often exploit two properties to reduce the problem size: \emph{dominance} and \emph{periodicity}.
Dominance relations are exploited to reduce~\(n\), and periodicity is exploited to reduce~\(c\).

\subsubsection{Simple, multiple, collective, and threshold dominances}

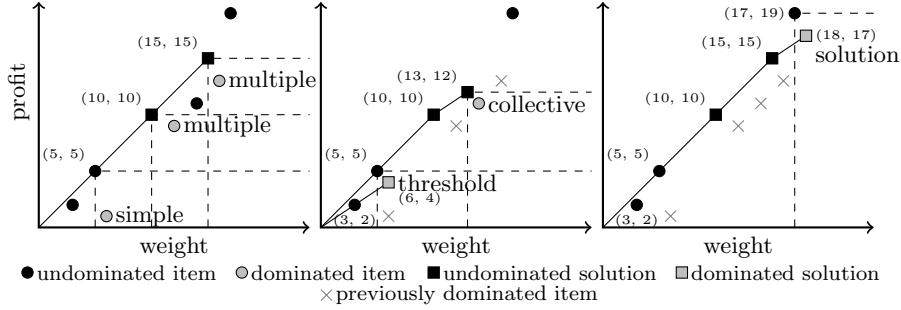
\begin{figure}
  \centering
  \caption{Examples of the simple, multiple, collective, threshold and solution dominances. The solution dominance is proposed by the authors in~\autoref{sec:oso_and_sol_dom}. The item set considered in the three graphs is: (3, 2), (5, 5), (6, 1), (12, 9), (14, 11), (16, 13), (17, 19). An item or solution dominates any item or solution that have the same weight or more and the same profit or less (dashed lines). The dominated items and solutions are labeled with the type(s) of dominance they are subject to. Source: primary.}
  \label{fig:dominances}
  \hspace{0.4cm}
\edef \scale {0.15}
\edef \myrad {0.5}
\begin{tikzpicture}[scale=\scale]

% simple & multiple dominance

\drawaxis{24}{20}{weight}{profit}

\draw [fill] (3, 2) circle [radius=\myrad];

\draw (0, 0) -- (15,15);
\draw [fill] (5, 5) circle [radius=\myrad];
\node [above left, font=\tiny] at (5, 5) {(5, 5)};
\draw [dashed] (5, 0) -- (5, 5) -- (24, 5);

\draw [dashed] (10, 0) -- (10, 10) -- (24, 10);
%\draw [fill=lightgray] (10 - \myrad, 10 - \myrad) rectangle +(2*\myrad, 2*\myrad);
\drawrectangle{10}{10}{\myrad}{black}
\node [above left, font=\tiny] at (10, 10) {(10, 10)};

\draw [dashed] (15, 0) -- (15, 15) -- (24, 15);
%\draw [blue,fill=white] (15, 15) circle [radius=\myrad];
\drawrectangle{15}{15}{\myrad}{black}
\node [above left, font=\tiny] at (15, 15) {(15, 15)};

%\draw [fill=red] (7, 2) circle [radius=\myrad];
%\node [right, font=\small] at (7, 2) {simple};
\draw [fill=lightgray] (6, 1) circle [radius=\myrad];
\node [right, font=\small, align=left] at (6, 1) {simple};
\draw [fill=lightgray] (12, 9) circle [radius=\myrad];
\node [right, font=\small] at (12, 9) {multiple};
\draw [fill] (14, 11) circle [radius=\myrad];
\draw [fill=lightgray] (16, 13) circle [radius=\myrad];
\node [right, font=\small] at (16, 13) {multiple};
\draw [fill] (17, 19) circle [radius=\myrad];

% collective & threshold dominance
\begin{scope}[shift={(25,0)}]
\drawaxis{24}{20}{weight}{}

\draw [] (0,0) -- (6,4);
\draw [dashed] (5, 0) -- (5, 5) -- (24, 5);
\draw [fill] (3, 2) circle [radius=\myrad];
\node [below, font=\tiny] at (3, 2) {(3, 2)};
\drawrectangle{6}{4}{\myrad}{lightgray}
\node [below right, font=\tiny] at (6, 4) {(6, 4)};
\node [right, font=\small] at (6,4) {threshold};

\draw [] (0,0) -- (10,10) -- (13,12);
\draw [fill] (5, 5) circle [radius=\myrad];
\node [above left, font=\tiny] at (5, 5) {(5, 5)};

%\draw [blue,fill=white] (10, 10) circle [radius=\myrad];
\drawrectangle{10}{10}{\myrad}{black}
\node [above left, font=\tiny] at (10, 10) {(10, 10)};

\draw [dashed] (13, 0) -- (13, 12) -- (24, 12);
%\draw [blue,fill=white] (13, 12) circle [radius=\myrad];
\drawrectangle{13}{12}{\myrad}{black}
\node [above left, font=\tiny] at (13, 12) {(13, 12)};

%\draw [fill] (7, 2) circle [radius=\myrad];
%\draw [fill] (11, 4) circle [radius=\myrad];
%\draw [fill=lightgray] (6, 1) circle [radius=\myrad];
\drawolddomitem{6}{1}
%\node [right, font=\small, align=left] at (6, 1) {simple};
%\draw [fill] (12, 9) circle [radius=\myrad];
\drawolddomitem{12}{9}
\draw [fill=lightgray] (14, 11) circle [radius=\myrad];
\node [right, font=\small] at (14, 11) {collective};
%\draw [fill] (16, 13) circle [radius=\myrad];
\drawolddomitem{16}{13}
\draw [fill] (17, 19) circle [radius=\myrad];
\end{scope}

% solution dominance
\begin{scope}[shift={(50 ,0)}]
\drawaxis{24}{20}{weight}{}

%\draw [dashed] (5, 0) -- (5, 5) -- (24, 5);
%\draw [] (0,0) -- (6,4);
\node [below, font=\tiny] at (3, 2) {(3, 2)};
\draw [fill] (3, 2) circle [radius=\myrad];
%\node [right, font=\tiny] at (6, 4) {(6, 4)};
%\draw [blue,fill=lightgray] (6, 4) circle [radius=\myrad];
%\drawrectangle{6}{4}{\myrad}{lightgray}
%\node [below right, font=\small] at (6,4) {threshold};

\draw [] (0, 0) -- (15,15) -- (18,17);

\draw [fill] (5, 5) circle [radius=\myrad];
\node [above left, font=\tiny] at (5, 5) {(5, 5)};

%\draw [blue,fill=white] (10, 10) circle [radius=\myrad];
\drawrectangle{10}{10}{\myrad}{black}
\node [above left, font=\tiny] at (10, 10) {(10, 10)};

%\draw [blue,fill=white] (15, 15) circle [radius=\myrad];
\drawrectangle{15}{15}{\myrad}{black}
\node [above left, font=\tiny] at (15, 15) {(15, 15)};

%\draw [blue,fill=red] (20, 20) circle [radius=\myrad];
%\node [above, font=\small] at (20, 20) {threshold};
%\node [right, font=\tiny] at (20, 20) {(20, 20)};

%\draw [fill] (7, 2) circle [radius=\myrad];
%\draw [fill=lightgray] (11, 4) circle [radius=\myrad];
%\node [right, font=\small, align=left] at (11, 4) {simple};
\drawolddomitem{6}{1}
\drawolddomitem{12}{9}
\drawolddomitem{14}{11}
\drawolddomitem{16}{13}
%\draw [fill] (14, 11) circle [radius=\myrad];

%\draw [blue,fill=lightgray] (18, 17) circle [radius=\myrad];
\drawrectangle{18}{17}{\myrad}{lightgray}
\node [right, font=\tiny] at (18,17) {(18, 17)};
\node [below right, font=\small] at (18,17) {solution};

\node [left, font=\tiny] at (17, 19) {(17, 19)};
\draw [fill] (17, 19) circle [radius=\myrad];
\draw [dashed] (17, 0) -- (17, 19) -- (24, 19);
\end{scope}

\draw [fill] (-1, -4) circle [radius=\myrad];
\node [right, font=\footnotesize] at (-1,-4) {undominated item};
\draw [fill=lightgray] (18, -4) circle [radius=\myrad];
\node [right, font=\footnotesize] at (18,-4) {dominated item};
\drawrectangle{35}{-4}{\myrad}{black}
\node [right, font=\footnotesize] at (35,-4) {undominated solution};
\drawrectangle{57}{-4}{\myrad}{lightgray}
\node [right, font=\footnotesize] at (57,-4) {dominated solution};
\drawolddomitem{25.5}{-6}
\node [right, font=\footnotesize] at (25.5,-6) {previously dominated item};

\end{tikzpicture}
\end{figure}

Any item~\(j\) that does not appear in all optimal solutions of an instance can be excluded without affecting our capability of solving such instance.
Given two items~\(i\) and~\(j : j \neq i\), if~\(w_i \leq w_j\) and~\(p_i \geq p_j\), then~\(j\) cannot be in all optimal solutions, since for any optimal solution that contains~\(j\) there will exist another optimal solution with~\(i\) in place of~\(j\).
Consequently, \(j\)~can be ignored when solving an instance that contains both~\(i\) and~\(j\).
Such relationship between~\(i\) and~\(j\) is called \emph{simple dominance}, more specifically we can say that~\(i\) simple dominates~\(j\) (or that~\(j\) is simple dominated by~\(i\)).
For an example, \((5, 5)\) simple dominates \((6, 1)\), as shown in \autoref{fig:dominances}.

Given a positive integer~\(\alpha\), if~\(\alpha \times w_i \leq w_j\) and~\(\alpha \times p_i \geq p_j\), then~\(\alpha\) copies of~\(i\) can replace one of~\(j\),  and~\(j\) can be ignored.
Such relationship is called \emph{multiple dominance}, and it generalizes simple dominance in which \(\alpha = 1\).
For an example, \(\alpha = 2\) copies of \((5, 5)\) multiple dominates \((12, 9)\).

Given a valid solution~\(s\), if~\(w_s \leq w_j\) and~\(p_s \geq p_j\), then the items that compose~\(s\) can replace~\(j\), and~\(j\) can be ignored.
Such relationship is called \emph{collective dominance}~\citep{ukp_new_results}, and it generalizes multiple dominance in which~\(s\) consists of~\(\alpha\) copies of~\(i\).
For an example, solution \(\{(5, 5), (5, 5), (3, 2)\}\) collective dominates \((14, 11)\).

If~\(w_s \leq \beta \times w_j\) and~\(p_s \geq \beta \times p_j\), then the items that compose~\(s\) can replace~\(\beta\) copies of~\(j\), and solutions including~\(\beta\) or more copies of~\(j\) can be ignored.
Such relationship is called \emph{threshold dominance}, and it generalizes collective dominance in which \(\beta = 1\).
Threshold dominance with~\(\beta > 1\) does not allow to exclude an item~\(j\) as a preprocessing phase, but it reduces the search space by allowing the algorithm to ignore all solutions in which~\(x_j \geq \beta\).
For an example, \((5, 5)\) threshold dominates \(\beta = 2\) copies of \((3, 2)\).
%The~\autoref{fig:dominances} presents a small set of items in which these four dominance relations can be observed.

\subsubsection{Periodicity}

\begin{figure}
  \centering
  \caption{
    Capacities and their connection with periodicity in an instance where \(y^+ < c\).
    % Algorithms that solve the UKP only for an specific capacity can solve for \(y^*\) instead of \(c\).
    % Algorithms that solve the UKP for all capacities up to \(c\) can solve up to \(y^*\) instead, but if they solve up to \(y^+\) then an optimal solution for any subsequent capacity can be computed in \(O(1)\).
    Source: primary.}
  \hspace{0.4cm}
  \label{fig:periodicity}
\edef \scale {0.5}
\begin{tikzpicture}[scale=\scale]
\drawhvectorfill{(0, 0)}{22}{1}{1}{white}
\node [scale=\scale, font=\Huge] at (0 + 0.5, 0.5) {\(0\)};
\node [scale=\scale, font=\Huge] at (1 + 0.5, 0.5) {\(1\)};
\node [scale=\scale, font=\Huge] at (2 + 0.5, 0.5) {\(2\)};
\node [scale=\scale, font=\Huge] at (3 + 0.5, 0.5) {\(3\)};
\node [scale=\scale, font=\Huge] at (4 + 0.5, 0.5) {\(...\)};
\node [scale=\scale, font=\Huge] at (8 + 0.5, 0.5) {\(y^*\)};
\node [scale=\scale, font=\Huge] at (10 + 0.5, 0.5) {\(y^+\)};
\node [scale=\scale, font=\Huge] at (20 + 0.5, 0.5) {\(c\)};
\node [scale=\scale, font=\Huge] at (21 + 0.5, 0.5) {\(...\)};
\node [scale=\scale, font=\Huge] at (22 + 0.5, 0.5) {\(\infty\)};
\draw[dashed, ultra thick, ->] (20.4, 1) to [out=135,in=45] (14.6, 1);
\node [scale=\scale, below, font=\Huge] at (17.4, 2) {\(-w_b\)};
\draw[dashed, ultra thick, ->] (14.4, 1) to [out=135,in=45] (8.6, 1);
\node [scale=\scale, below, font=\Huge] at (11.4, 2) {\(-w_b\)};
%\draw[dashed, ultra thick, ->] (12.4, 1) to [out=135,in=45] (8.6, 1);
%\node [scale=\scale, above, font=\Huge] at (10.5, 2) {\(-w_b\)};
\draw[decorate, decoration={brace, mirror, raise=5pt}] (0.1, 0) -- (4.9, 0);
\node [below, font=\small, align=center] at (2.5, -0.5) {not relevant for \\ periodicity};
\draw[decorate, decoration={brace, mirror, raise=5pt}] (5.1, 0) -- (10.9, 0);
\node [below, font=\small, align=center] at (8, -0.5) {\([y^+ + 1 - w_{b}, y^+]\)};
\draw[decorate, decoration={brace, mirror, raise=5pt}] (11.1, 0) -- (22.9, 0);
\node [below, font=\small, align=center] at (17, -0.5)
{ optimal solutions can be computed \\
  in \(O(1)\) if optimal solutions for range \\
  \([y^+ + 1 - w_{b}, y^+]\) are known};
\end{tikzpicture}
\end{figure}
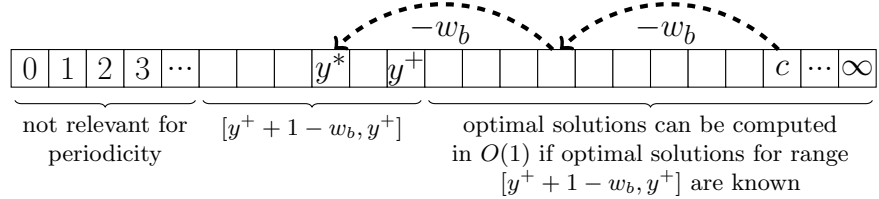

%After such capacity, all items but the best one are dominated.
%In other words, for any capacity greater than~\(y^+\), an optimal solution can be reached by adding copies of \(b\) to an optimal solution of capacity~\(y^+\) or lower.
The \emph{periodicity} property shows the existence of a capacity \(y^+\) such that, for every capacity~\(y'\)~:~\(y'>y^+\), there exists an optimal solution for capacity~\(y'\) that is the same as an optimal solution for capacity~\(y' - w_b\) except it includes one more copy of the best item~\(b\)~\citep[p.~10]{gg-66}.
If~\(y^+ < c\), then the UKP can be solved for capacity~\(y^{*} = c - \ceil{(c - y^+)/w_b}w_b\), and the gap between~\(y^{*}\) and~\(c\) filled with exactly~\((c - y^{*})/w_b\) copies of the best item \(b\), effectively reducing the knapsack size from~\(c\) to~\(y^*\) (see~\autoref{fig:periodicity}).
However, the effort necessary to compute the exact value of \(y^+\) (and \(y^*\)) is about the same as solving the UKP for all capacities \(y \leq y^+\) while checking for threshold dominance.
Consequently, \(y^+\) is implicitly determined by some dynamic programming algorithms (as EDUK) while solving instances in which \(y^+ < c\).
An upper bound on~\(y^+\) is less valuable than~\(y^+\) itself, but it can be computed in polynomial time, as a preprocessing phase.
The authors discuss the usefulness of computing or bounding~\(y^+\) in~\autoref{sec:critique_of_periodicity}.

The periodicity property is a direct consequence of the threshold dominance.
Except by the best item~\(b\), every item~\(j : j \neq b\) is threshold dominated for some~\(\alpha_j\).
For example, given that~\(y\) is the lowest common multiple of~\(w_b\) and~\(w_j\),~\(\beta_j = y/w_b\) and~\(\alpha_j = y/w_j\), then~\(\beta_j \times w_b \leq \alpha_j \times w_j \equiv y \leq y\) and~\(\beta_j \times p_b \geq \alpha_j \times p_j\). %, since \(e_b \geq e_j\) (from the definition of the best item).
Consequently, the threshold dominance defines a constraint~\(x_j < \alpha_j\) on every non-best item~\(j\).
All solutions that do not break such constraints and do not make use of the best item weight less than~\(y'' = \sum \alpha_j \times w_j\).
Consequently, an optimal solution for any capacity greater than \(y''\) can be obtained by the procedure described in the previous paragraph.
The capacity~\(y''\) is an upper bound on~\(y^+\)~\citep[p.~215]{book_ukp_2004}.

\section{The revisited ordered step-off and (partial) solution dominance}
\label{sec:oso_and_sol_dom}

The ordered step-off (OSO) is a dynamic programming algorithm for the UKP~\citep{gg-66}.
The OSO is described in Algorithm 1, a complementary overview follows.
The arrays \(g\) and \(d\) store, respectively, the profit value of generated solutions and the index of the most efficient item present in such solutions.
Together they can be seen as a solution pool indexed by solution weight.
A summary of the algorithm follows: the solution pool is initialized with all single-item solutions; the solution pool is iterated by weight order; for each solution, the solution pool is expanded with new solutions, each new solution is a copy of the current solution plus an extra item.
This process enumerates all undominated solutions, and it returns the optimal profit value.
The algorithm skips the creation of new solutions from old solutions already known to be dominated (lines~\ref{if_less_than_opt_begin}~to~\ref{if_less_than_opt_end}), discards some dominated solutions created (e.g., if two or more solutions share the same weight, the algorithm keeps only one of these which is tied for the highest profit), and avoids considering symmetric solutions (by restricting the loops up to \(d[y]\), the algorithm only considers the permutation of the solution in which the items are added in order of efficiency).
When the algorithm finishes executing, \(opt\) contains the optimal profit value and for every \(g[y] > 0\) there is a solution with: weight~\(y\), profit value \(g[y]\), and the index of the last (and most efficient) item added~\(d[y]\).

The authors revisited OSO and added the \emph{else if} found in lines \ref{added_elsif}~to~\ref{if_new_lower_bound_end} of Algorithm 1 (to distinguish this version from the original one, it will be called Revisited OSO, or R-OSO).
When two or more solutions generated have the same weight and profit (a special case of solution dominance), the original algorithm arbitrarily keeps the first solution generated and ignores the others.
In the same situation, the revisited algorithm keeps the solution that will generate less descendents (smallest \(d[y]\)), therefore minimizing the computational effort.
This change of tiebreaker reduced the run times of the algorithm over subset-sum and strongly correlated instances by orders of magnitude.
Subset-sum instances are UKP instances with items respecting \(\forall i.~p_i = w_i\), while strongly correlated instances have items respecting \(\forall i.~p_i = w_i + \alpha\) (where \(\alpha\) is a small positive integer value which is the same for the whole instance).
Therefore, the performance gain can be explained by the fact that, in subset-sum instances, all solutions with the same weight have the same profit and, in strongly correlated instances, all solutions with the same weight and the same number of items have the same profit.
%Without this change, between all solutions of the same weight, the algorithm would keep the first solution generated with the greatest profit value.
%With the change, between all solutions with the same weight, the algorithm keeps the first solution generated with the greatest profit value and smallest index of the last item added to the solution.

\begin{algorithm}[!htb]
\caption{The revisited ordered step-off (R-OSO)}
\begin{algorithmic}[1]
\Procedure{oso}{$n, c, w, p$}
  \State~Sort \(w\) and \(p\) by non-increasing item efficiency\footnotemark and find $w_{min}, w_{max}$ 
  \State~\(g \gets\) array of profit values with size~\(c+1\), initialized with zeroes\label{create_g}
  \State~\(d \gets\) array of item indexes with size~\(c+1\), values uninintialized\label{create_d}
  
  \For{\(i \gets n\) to \(1\)}\label{begin_trivial_bounds}\Comment{Stores one-item solutions}
%  \For{\(i \gets 1\) to \(n\)}\label{begin_trivial_bounds}\Comment{Stores one-item solutions}
%    \If{\(g[w_i] < p_i\)}
      \State~\(g[w_i] \gets p_i\)
      \State~\(d[w_i] \gets i\)
%    \EndIf
  \EndFor\label{end_trivial_bounds}

  \State~\(opt \gets 0\)\label{init_opt}

  %\For{\(y \gets w_{min}\) to \(c - w_{min}\)}\label{main_ext_loop_begin}%\Comment{Can end earlier because of periodicity check}
  \For{\(y \gets w_{min}\) to \(c\)}\label{main_ext_loop_begin}%\Comment{Can end earlier because of periodicity check}
    \If{\(g[y] \leq opt\)}\label{if_less_than_opt_begin}\Comment{Prunes dominated solutions}
        \State \textbf{continue}\label{alg:continue}\Comment{Ends current iteration and begins the next}
    \EndIf\label{if_less_than_opt_end}
    
    \State~\(opt \gets g[y]\)\label{update_opt}
    
    \For{\(i=1\) to \(d[y]\)}\label{main_inner_loop_begin}\Comment{Creates new solutions (never symmetric)}
      \If{\(y + w_i > c\)}\Comment{To avoid accessing invalid positions}
        \State~\textbf{continue}
      \EndIf
      \If{\(g[y + w_i] < g[y] + p_i\)}\label{if_new_lower_bound_begin}
        \State~\(g[y + w_i] \gets g[y] + p_i\)
        \State~\(d[y + w_i] \gets i\)
      \ElsIf{\(g[y + w_i] = g[y] + p_i\) \textbf{and} \(i < d[y + w_i]\)}\label{added_elsif}
        \State~\(d[y + w_i] \gets i\)
      \EndIf\label{if_new_lower_bound_end}
    \EndFor\label{main_inner_loop_end}
  \EndFor\label{main_ext_loop_end}

%  \For{\(y \gets c - w_{min} + 1\) to \(c\)}
%    \If{\(g[y] > opt\)}
%      \State~\(opt \gets g[y]\)
%    \EndIf
%  \EndFor
  \State \textbf{return}~\(opt\)

%  \For{\(y \gets c-w_{min}+1, c\)}\label{get_y_opt_loop_begin}\Comment{Removal of dominated solutions}
%    \If{\(g[y] > opt\)}\label{last_loop_inner_if}
%      \State~\(opt \gets g[y]\)
%      \State~\(y_{opt} \gets y\)
%    \EndIf
%  \EndFor\label{get_y_opt_loop_end}
\EndProcedure
\end{algorithmic}
\end{algorithm}
\footnotetext{If two or more items share the same efficiency they are sorted by non-decreasing weight.}

Given two valid solutions~\(s\) and~\(t\), if~\(w_s \leq w_t\) and~\(p_s \geq p_t\), then the items of solution~\(s\) can replace the items of solution~\(t\), and any solution that is a superset of \(t\) can be ignored (see \autoref{fig:dominances}).
The authors call this relationship \emph{solution dominance}, and it generalizes threshold dominance in which solution~\(t\) can only consist of copies of the same item.
As far the authors know, the concept of solution dominance was not formalized before.
The authors believe such overlook happened because: (1) the original concept of dominance focused on discarding single items from further consideration; (2) an algorithm that fully exploits such dominance relation has not been proposed.

The authors also propose the term \emph{partial solution dominance} to describe the mechanism by which (R-)OSO avoids the generation of some dominated solutions (but not all of them) with no overhead.
The mechanism consists in skipping the generation of up to~\(n\) child solutions (and all their descendents) because the original solution is already known to be dominated at the time they will be generated (lines~\ref{if_less_than_opt_begin}~to~\ref{if_less_than_opt_end}), combined with the fact no solution is enumerated two times (no symmetric solution thanks to the use of~\(d[y]\)).
Our experiments show that, while \emph{partial solution dominance} lacks the same guarantees that applying simple, multiple, collective, and threshold dominances as soon as possible (as in EDUK), its performance is state of the art.

A formal description of which dominated solutions are guaranteed to not be generated and which ones may yet be generated follows.
The notation~\(min_{ix}(s)\) refer to the index of the item of lowest index present in solution~\(s\); the notation~\(max_{ix}(s)\) has the analogue meaning.
If a solution~\(t\) is skipped because solution~\(s\) dominates~\(t\) (lines~\ref{if_less_than_opt_begin} to~\ref{if_less_than_opt_end}), only the solutions~\(u\) where \(t \subset u\) and \(max_{ix}(u - t) \leq min_{ix}(t)\) are guaranteed to not be generated anymore.
However, any solutions~\(t \cup \{i\} : i > min_{ix}(t)\) yet generated are dominated by~\(s \cup \{i\}\) (or by a solution that dominates~\(s \cup \{i\}\)).
Consequently, such solutions are guaranteed to be skipped and the amount of supersets of~\(t\) generated is further reduced.
Unfortunately, after skipping those solutions, the solutions~\(t \cup \{i, j\} : j > i > min_{ix}(t)\) may be yet generated, and so on.

For an example, if~\(\{3, 2\}\) is dominated, then~\(\{3, 2, 2\}\) and \(\{3, 2, 1\}\) are not generated by OSO and, consequently, \(\{3, 2, 2, 2\}, \{3, 2, 2, 1\}\) and \(\{3, 2, 1, 1\}\) are also not generated, and so on.
However, given~\(\{4, 3\}\) is not dominated, it will be used to generate~\(\{4, 3, 2\}\) (which is a superset of \(\{3, 2\}, \{4, 2\}\) and \(\{4, 3\}\)), and is dominated as~\(\{3, 2\}\) is dominated.
Ideally, \(\{4, 3, 2\}\) should not be generated, but: (1) if the weight of~\(\{4, 3\}\) is lower than~\(\{3, 2\}\), then~\(\{3, 2\}\) is not even known to be dominated at the time~\(\{4, 3, 2\}\) is generated; (2) if the weight of~\(\{4, 3\}\) is greater than~\(\{3, 2\}\), yet we would need to check all subsets of~\(\{4, 3, 2\}\) with one less item, this is a~\(O(n)\) overhead, greater than letting the dominated solution be generated, and skipping it after.

The (R-)OSO partial solution dominance mechanism is rendered useless if the instance has some properties.
However, such properties seem to not be found in instances of the literature, nor seem to arise naturally in the real-world or in current artificial item distributions.
The properties are: the same item has the lowest weight and the lowest efficiency; the second lowest weight is orders of magnitude larger than the lowest weight.
In the most extreme cases, this is, when the least efficient item has one unit of weight and the second lowest weight is close to~\(c\), the (R-)OSO will display its worst-case behavior, reverting to the performance of the naïve DP algorithm (\(nc\) steps).
This happens because: there will be a solution for each capacity (i.e., no sparsity); the least efficient item has index~\(n\) and each solution visited before the second lowest weight will, therefore, generate~\(n\) new solutions (many possibly discarded for being larger than the knapsack).

\section{A critique of periodicity checks and bounds}
\label{sec:critique_of_periodicity}

%Some prior work presents periodicity checks and bounds as ways of improving the performance of exact algorithms for the UKP.
%However, such techniques contribute almost nothing to the performance of the current algorithms for the UKP.
%The purpose of this section is to examine this often overlooked situation.

State-of-the-art dynamic programming algorithms like EDUK~\citep{eduk} and the terminating step-off~\citep{gg-66} finish the execution early if they detect that only the best item will be used to generate new solutions.
This mechanism is called \emph{periodicity check}.
If a periodicity check finishes the execution at capacity~\(y\), it saves the effort of iterating the last \(c - y\) capacities.
However, as only the best item is yet used, the algorithm would execute only \(\Theta(1)\) operations for each capacity in the range \([y+1,c]\).
Periodicity checks often demand about \(n\) and up to \(y \leq c\) extra operations to save up to \(c - y\) operations.
Consequently, periodicity checks save little effort if any, in comparison to the application of dominances which they depend on to work.

If periodicity checks save little effort, then the computation of upper bounds on~\(y^+\) save up even less effort.
When the global item list of EDUK is reduced to the best item, the current capacity is probably a far better upper bound on~\(y^+\) than any upper bound computed by a polynomial algorithm.
While branch-and-bound algorithms do not have periodicity checking, they also do not obtain any considerable benefit from periodicity bounds, as their bound computation already supersedes them.
Consequently, periodicity checks and bounds are of little relevance for the performance of state-of-the-art algorithms for the UKP.

%In a dynamic programming algorithm with periodicity checking, the utility of upper bounds on~\(y^+\) is restricted to saving memory and saving the time spent initializing it (which is insignificant).

%If an upper bound on \(y^+\) can be computed, item dominance and the periodicity checks stops the algorithm before the capacity predicted by the upper bound.

\cite{pya} states that if~\(w_{max} \leq c/2\) then ``computing all optimal states \((y,f(N,y))\) with~\(y \leq c/2\) is enough\footnotemark[5], since any knapsack with capacity \(y \in~] c/2 , c]\) can be solved by completing the solution of~\(\text{UKP}^{\text{\smash{\(y - c/2\)}}}_{\text{\textbf{w, b}}}\) with the one of~\(\text{UKP}^{\text{\smash{\(c/2\)}}}_{\text{\textbf{w, b}}}\).''.
If this claim was correct and~\(w_{max} \leq c/2^x : x \geq 1\), then a dynamic programming algorithm for the UKP would never need to solve for a capacity greater than or equal to~\(2 \times w_{max}\).
It would suffice to solve the instance up to capacity~\(c/2^x\) and then multiply the number of items of each type in an optimal solution by~\(2^x\).
Unfortunately, the claim is incorrect.
A counterexample is presented below.
Consider the following instance: \(c = 6\), \(n = 2\), \(w_1 = p_1 = 1\), \(w_2 = 2\), \(p_2 = 10\).
The optimal solution for capacity~\(c/2 = 3\) is~\(11\) (one copy of each item), but the optimal solution for~\(c\) is~\(30\) (three copies of item two) not~\(22\) (two copies of each item).
\footnotetext[5]{This test was not implemented in EDUK.}

%While on the topic of reducing \(c\), the authors would like to present a counterexample for the following statement: ``computing all optimal states \((y,f(N,y))\) with \(y \leq c/2\) is enough\footnotemark[5], since any knapsack with capacity \(y \in~] c/2 , c]\) can be solved by completing the solution of \(\text{UKP}^{y - c/2}_{\text{\textbf{w, b}}}\) with the one of \(\text{UKP}^{c/2}_{\text{\textbf{w, b}}}\).''~\cite{pya}. Consider the following instance: \(c = 6\), \(n = 2\), \(w_1 = p_1 = 1\), \(w_2 = 2\), \(p_2 = 10\). The optimal solution for \(c/2 = 3\) is \(11\) (one copy of each item), but the optimal solution for \(c\) is \(30\) (three copies of item two) not \(22\) (two copies of each item).
%\footnotetext[5]{This test was not implemented in EDUK.}

\section{Methods}
\label{sec:methods}

This section describes the algorithms, their implementations, the instance datasets and the computer setup used.
The authors also present a rationale for not including some algorithms and datasets from the literature.

\subsection{Algorithms and their implementations}

%The algorithms used in the experiments belong to three main approaches: dynamic programming (DP); branch-and-bound (B\&B); and hybrid (combine DP and B\&B).
The usual dynamic programming (DP) algorithms for the UKP have a worst-case time complexity of~\(O(nc)\) (pseudo-polynomial), and worst-case space complexity of~\(O(n+c)\).
The branch-and-bound (B\&B) algorithms have an exponential worst-case time complexity (all item combinations that fit the knapsack), 
but the worst-case space complexity of the B\&B algorithms used in the experiments is~\(O(n)\).
%The hybrid algorithm used in the experiments displays the same time and space worst-case complexities than the DP approach.

% B&B INFO TO ADD: branching scheme, bounds used commonly, ...

MTU1 is a B\&B algorithm for the UKP~\citep{mtu1}. %with worst-case space complexity \(O(n)\)\cite{mtu1}.
In the enumeration tree of MTU1, each node at depth~\(0 \leq d < n\) has~\((c'/w_{d+1}) + 1\) children (where \(c'\) is the remaining capacity at the current node), with each child representing a valid amount of copies of the item~\(d+1\) being packed in the knapsack (including zero copies).
In practice, since it follows a depth-first search, MTU1 has no explicit tree, but instead, make changes directly over the current solution, which represents the current branch from the root to a leaf.
Items are added to and removed from the solution to simulate the tree traversal.
The exploration order favors the siblings representing higher amounts before lower amounts.
As the items are sorted by efficiency, this means the first node after the root will represent the maximum amount of copies of the best item, and the path to the first visited leaf will be the solution given by the classic greedy heuristic which packs the most efficient item that yet fits the knapsack until there is no item that fits the knapsack. Consequently, the algorithm already begins with this lower bound.
The upper bound computed by MTU1 in each node is the continuous relaxation of the problem with the already set variables fixed, which is solved by multiplying the efficiency of the most efficient item not yet fixed by the remaining capacity.

% DP INFO TO ADD: states, stages, recursive function

The MTU2 algorithm calls MTU1 over the \(x\%\) most efficient items, and if this is not sufficient to obtain a solution with proved optimality, it repeats the process with \(x\%\) increased~\citep{mtu2}.
The Fortran implementation of MTU1 and MTU2 used in our experiments was the original implementation by Martello and Toth but with all 32 bits integers or float variables/parameters replaced by their 64 bits counterparts. This version is publicly available in the authors' code repository\footnote{The MTU1 and MTU2 adapted Fortran code used in the experiments is available at \url{https://github.com/henriquebecker91/masters/tree/136c1c1fbeb6ef7baa7ab6bcc8f48cb0bb68b697/codes}}.
The C++ implementations of both algorithms were written by the authors.
The C++ and Fortran implementations of the MTU1 algorithm have no significant differences.

The MTU2 implementations differed in the algorithm used to partially sort the items and the exact ordering.
\cite{mtu2} does not specify the exact method for performing the partial sorting.
The original Fortran implementation uses a complex algorithm developed by one of the authors of MTU2 in~\cite{partial_sort_martello} to find the k\textsuperscript{th} most efficient item in an unsorted array, and then select and sort only the items that have the same or a greater efficiency.
The C++ implementation uses the \verb+std::partial_sort+ procedure of the standard C++ library \verb+algorithm+.
The Fortran implementation only sorts the items by nonincreasing efficiency; the C++ implementation breaks efficiency ties sorting by nondecreasing weight.

A description of the recursion, states, and stages in which the DPs for the UKP are based follow.
Given \(opt(y)\) denotes the optimal solution value for capacity~\(y\), the recursion for the UKP can be written as \(opt(y) = max \{0, p_1 + opt(y - w_1), p_2 + opt(y - w_2), \dots, p_n + opt(y - w_n)\}\) (where \(\forall~y < 0.~opt(y) = -\infty\)).
The UKP has a single state, that is the remaining knapsack capacity~\(y\). 
Different from 0-1 KP and BKP, the UKP has no need to take into account which items were already used at each decision as there is an unlimited amount of copies available for each item.
This difference allows UKP to need only \(O(c)\) memory, instead of \(O(nc)\) as in 0-1 KP and BKP.
For each capacity~\(y = c - w_s\) (where \(s\) is a valid solution), there is a decision point; consequently, the number of stages is not exact but can go up to~\(c\) (as in the case of \(w_{min} = 1\)).

The ordered step-off (OSO) is a DP algorithm proposed in~\cite[p.~15]{gg-66}.
The authors already presented and discussed a revisited version of OSO in~\autoref{sec:oso_and_sol_dom}.
The terminating step-off (TSO) is the same as OSO but it includes periodicity checking.
%which executes \(\theta(n) + O(c)\) extra operations to save \(O(c)\) operations.
\cite{green_improv} proposes another variant of OSO, referred in this paper as GFDP (Greenberg and Feldman's Dynamic Programming).
The GFDP does not use the best item~\(b\), but interrupts the DP at each~\(w_b\) capacity positions to verify if the DP can stop, and the remaining capacity filled with copies of \(b\). 
If two or more items share the greatest efficiency, GFDP verification does not work, and it is the same as OSO. 

The original implementations of the step-offs and GFDP were not publicly available, so the authors wrote their own implementations in C++.
The authors' implementations of TSO and GFDP include the same tiebreaking improvement added by the authors to OSO and described in~\autoref{sec:oso_and_sol_dom}.
All C++ implementations written by the authors are available at the first author's code repository\footnote{The C++ implementations of MTU1, MTU2, the revisited ordered/terminating step-offs, and R-GFDP are available at \url{https://github.com/henriquebecker91/masters/tree/136c1c1fbeb6ef7baa7ab6bcc8f48cb0bb68b697/codes/cpp}.}.
%As the C++ implementations of MTU1 and MTU2, our implementations change the items sorting by nonincreasing efficiency (specified by these three algorithms) to sorting by nonincreasing efficiency with efficiency ties sorted by nondecreasing weight.

EDUK (Efficient Dynamic programming for the Unbounded Knapsack problem) was the first DP algorithm to explicitly check for threshold dominance (a concept proposed together with the algorithm) and collective dominance (that was independently discovered by Pisinger~\citep{pisinger1994dominance}), it also features a sparse representation of the iteration domain~\citep[p.~223]{ukp_new_results,eduk,book_ukp_2004}.
EDUK seems to be based on ideas first discussed in~\cite{algo_tech_cut}.
Before EDUK2 was proposed, it was said that ``[...] EDUK [...] seems to be the most efficient dynamic programming based method available at the moment.''~\citep[p.~223]{book_ukp_2004}.

EDUK2 is a hybrid DP/B\&B algorithm which improves EDUK with a B\&B preprocessing phase~\citep{pya}.
If B\&B can solve the instance using less than a parameterized number of nodes, then EDUK is not executed; otherwise, the bounds computed in the B\&B phase are used to reduce the number of items before EDUK execution and in intervals during its execution.

The implementation of EDUK and EDUK2 used in the experiments was PYAsUKP (PYAsUKP: Yet Another solver for the Unbounded Knapsack Problem), which was written in OCaml.
Vincent Poirriez gave access to this code to the authors, by email,
in January 11th, 2016\footnote{The code is available at Henrique Becker master's thesis code repository (\url{https://github.com/henriquebecker91/masters/blob/f5bbabf47d6852816615315c8839d3f74013af5f/codes/ocaml/pyasukp_mail.tgz}).}.
%The version of the code available at that time in the PYAsUKP official site had bugs\footnote{The EDUK and EDUK2 source was available to download in the following page of PYAsUKP official site: \url{http://download.gna.org/pyasukp/pyasukpsrc.html}.}.

Finally, the authors also implemented the UKP formulation (i.e., Eq. \eqref{eq:objfun}--\eqref{eq:x_integer}) using the C++ interface of both Gurobi~8.0.1~\citep{gurobi} and CPLEX~12.8~\citep{cplex} to solve single UKP instances.
To make comparison to other methods fair and the results reproducible, the solvers were configured to: execute in single thread and deterministically (i.e., random seed fixed to zero); finish before time limit only with a \(0\%\) gap between the upper and lower bounds (i.e., search for a proven optimal solution); have the smallest tolerance possible to variable values deviating from integrality (without this about \(9\%\) of the results were slightly below or above the optimal value)\footnote{The exact parameters used for each solver were: GRB\_IntParam\_Threads (Gurobi) and Threads (CPLEX); GRB\_IntParam\_Seed (Gurobi) and RandomSeed (CPLEX); GRB\_DoubleParam\_MIPGap (Gurobi) and MIP::Tolerances::MIPGap (CPLEX); GRB\_DoubleParam\_IntFeasTol (Gurobi) and MIP::Tolerances::Integrality (CPLEX). All CPLEX parameters are prefixed by IloCplex::Param. The codes are available in: \url{https://github.com/henriquebecker91/masters/tree/efea8a981a72237edd17fedb4742ac568ded831c/codes/cpp/lib} (files \emph{cplex\_ukp\_model.hpp} and \emph{gurobi\_ukp\_model.hpp}). }.

\subsubsection{Algorithms deliberately ignored}

%Both the naïve DP algorithm for the UKP~\cite[p.~311]{tchu} and an improved version of it presented in~\cite[p.~221]{garfinkel} are dominated by OSO algorithm presented in~\cite[p.~15]{gg-66}.
The naïve DP algorithm for the UKP~\cite[p.~311]{tchu}, an improved version of it presented in~\cite[p.~221]{garfinkel} and the OSO~\cite[p.~15]{gg-66} are all \(O(nc)\) DP algorithms similar to each other. % These three DP algorithms are \(O(nc)\).
However, OSO does not need to execute \(n\) operations for each distinct \(c\) value and, in practice, will iterate only a small fraction of \(n\) (or even an empty list) for most \(c\) values of most instances.
The other two algorithms \emph{always} execute \(nc\) operations regardless of any instance properties. 
Preliminary tests confirmed that OSO dominated the other two algorithms and, consequently, both were not included in our experiments.
% Consequently, both the naïve DP algorithm and its improved version were not included in our experiments.

The UKP5 algorithm proposed in~\cite{sea2016} was found to be very similar to TSO\footnote{
  The authors of this article reinvented an algorithm from~\cite{gg-66} and published a paper calling it UKP5 while believing it was novel~\cite{sea2016}.
  The authors would like to apologize to the scientific community for such disregard.
  The only improvement of UKP5 over TSO is the tiebreaker change described in \autoref{sec:oso_and_sol_dom}, which the authors included in all algorithms it was applicable.}
and therefore only TSO was included.

The authors' implementation of the first algorithm proposed in~\cite{on_equivalent_greenberg} exceeded the time limits we used in the experiments, while the second algorithm does not work for all UKP instances\footnote{The authors' implementations of both algorithms were made available at \url{https://github.com/henriquebecker91/masters/blob/e2ff269998576cb69b8d6fb1de59fa5d3ce02852/codes/cpp/lib/greendp.hpp}.}.
Both weren't included in the experiments.
The Sage-3D algorithm from \cite{landa_sage} cited in \cite{ukp_hu_landa_shing_survey} needs \(O(n w_b p_b)\) memory and time, which is prohibitive for many instances considered and, therefore, was also not included.
The algorithm holds a considerable theoretical value, and its complexity is justified by the fact Sage-3D does not solve the UKP for a specific knapsack capacity, but instead, builds a data structure which allows querying the solution for a specific capacity in \(O(log(p_b))\).

In~\cite{mtu1}, the B\&B algorithm proposed in~\cite{gg-63} is said to be two times slower than the algorithm proposed in~\cite{cabot}, which was found to be dominated by MTU1; 
also, the algorithm in~\cite{gg-63} seems to have been abandoned by its authors in favor of OSO. %Also, \cite{cabot}~was behind a paywall.
Thus, the B\&B algorithms proposed in~\cite{gg-63} and~\cite{cabot} were not included in the experiments.

%Likewise, \cite{turnpike}~is behind a paywall, and 
In~\cite{green_improv} it is implied that GFDP is an improved version of the algorithm proposed in~\cite{turnpike}, so only GFDP was included.
The authors could not obtain the code of the algorithm proposed in~\cite{babayev}, and they chose to not reimplement it to not risk misrepresenting it, as it was not trivial to implement.

UKP-specific algorithms perform better than applying BKP or 0-1 KP algorithms over converted UKP instances~\cite{mtu1}, so BKP and 0-1 KP algorithms were not considered.

\subsection{Instance datasets}

The datasets used in the experiments include: artificial datasets from the literature that focus on being hard-to-solve (PYAsUKP and realistic random datasets); a dataset proposed by the authors in order to prove an hypothesis (BREQ dataset); a dataset based on solving of CSP/BPP instances with the column generation technique (CSP dataset).
The reasoning for not including \emph{uncorrelated} and \emph{weakly correlated} instances is presented in the end of the section.

\subsubsection{PYAsUKP dataset}
\label{sec:pya_inst}

The PYAsUKP dataset is described in~\cite{sea2016}, and comprises 4540 instances from five smaller datasets.
The PYAsUKP dataset was heavily based on the datasets presented in~\cite{pya}, which were used to compare EDUK2 to other UKP solving algorithms.
The instance generator used to generate the 4540 instances share the code with EDUK/EDUK2 implementations (PYAsUKP), which is the reason we call this dataset the \emph{PYAsUKP dataset}.
The PYAsUKP dataset comprises: 400 subset-sum instances (\(10^3 \leq n \leq 10^4\)); 240 strongly correlated instances (\(5\times10^3 \leq n \leq n = 10^4\)); 800 instances with postponed periodicity (\(2\times10^4 \leq n \leq 5\times10^4\)); 2000 instances without collective dominance (\(5\times10^3 \leq n \leq 5\times10^4\)); 1100 SAW instances (\(10^4 \leq n \leq 10^5\)).
The PYAsUKP dataset has multiple-of-ten amounts of instances generated with different random seeds for each combination of the remaining generation parameters (\(n\), \(w_{min}\), \dots).
The authors selected the first one-tenth of the instances for each parameter combination\footnote{The entire PYAsUKP dataset is available at \url{https://drive.google.com/open?id=0B30vAxj_5eaFSUNHQk53NmFXbkE}. The instances with the same parameter combination are numbered.} (in total 454) and will refer to it as the \emph{reduced PYAsUKP dataset}.

\subsubsection{Realistic Random Dataset}
A dataset of \emph{realistic random} instances was used in the experiments. The generation procedure, based on~\cite{eduk}, is summarized as follows: generate two lists of~\(n\) unique random integers uniformly distributed in~\([min, max]\) and sort them by increasing value; combine both lists in an item list, by pairing up the i-th of one list to the i-th element of the other; randomly shuffle the item list; generate a random capacity~\(c \in [c_{min},c_{max}]\) (uniform distribution).
Simple dominance cannot occur in such instances; other dominances may be present.
Our dataset comprises ten instances generated with distinct random seeds for each one of eight \(n\) values (\(2^{n'}\), where \(n' \in \{10, 11, \dots, 17\}\)), totalling 80 instances.
The values of the remaining parameters come from \(n\): \(max = n \times 2^{10}\), \(min = max/2^4\), \(c_{min} = 2\times max\), and \(c_{max} = c_{min} + min\).

\subsubsection{BREQ 128-16 Standard Benchmark}
\label{sec:breq_inst}

The Bottom Right Ellipse Quadrant (BREQ) is an items distribution proposed in~\cite{ukp_hb_mastersthesis}. %and first described in REF\_MASTER\_THESIS.
The items of an instance follow the BREQ distribution iff the profits and weights respect~\(p_i = p_{max} - \floor[\big]{\sqrt{p_{max}^2 - w_i^2 \times (p_{max}/w_{max})^2}}\), where \(w_{max}\) (\(p_{max}\)) is an upper bound on the items weight (profit).
The distribution name comes from the fact that the formula describes the bottom right quarter of an ellipse.
%This instance distribution was created to illustrate that different item distributions favor different solution approaches and, therefore, the choice of instances (or specifically, their item distribution) defines what is considered the \emph{best algorithm}.

The purpose of this items distribution is to illustrate the authors' point that artificial distributions can be developed to favor one solving approach over another.
In the case of the BREQ distribution, it favors B\&B over DP.
Distributions with the opposite property (favor DP over B\&B) are common in the recent literature.

The optimal solution of BREQ instances is often in the first fraction of the search space examined by B\&B algorithms. 
Moreover, the lower bounds from good solutions allow B\&B methods to skip a large fraction of the search space and promptly prove optimality.
In BREQ instances, the presence of simple, multiple and collective dominance is minimal
\footnote{
If the BREQ formula did not include the rounding, the profit of the item would be a strictly monotonically increasing function of the items weight.
Any item distribution with this property cannot present simple, multiple, or collective dominance. 
}, but threshold dominance is widespread: an optimal solution will never include the item~\(i\) two or more times if there is an item~\(j\) such as that~\(\sqrt{2} \times w_i \leq w_j \leq 2 \times w_i\).
Such characteristics lead to optimal solutions comprised of the largest weight items, which do not reuse optimal solutions for lower capacities.
This means that solving the UKP for lower capacities as DP algorithms do is mostly a wasted effort.

%If those three dominance relations are absent, for any solution~\(s\) composed of two or more items, and for any single item~\(i\), if~\(w_s \leq w_i\) then~\(p_s < p_i\).

% Proof that this interval is tight: http://www.wolframalpha.com/input/?i=2*(100+-+sqrt(100%5E2+-+w%5E2+*+16%5E2))+%3C%3D+100+-+sqrt(100%5E2+-+((sqrt(2)*w)%5E2+*+16%5E2))

The authors named the BREQ dataset used in the experiments of \emph{BREQ 128-16 Standard Benchmark}.
This dataset comprises 100 instances generated from all combinations of ten random seeds and ten distinct \(n\) values  defined as \(n = 2^{n'}\), where \(n' \in \{11, 12, \dots, 20\}\).
The values of the remaining parameters can be computed as follows: \(p_{min} = w_{min} = 1\), \(c = 128 \times n\), \(w_{max} = c\) and \(p_{max} = 16 \times w_{max}\).
The items generation procedure follows:
generate~\(n\) unique random integer weights uniformly distributed in~\([w_{min},w_{max}]\);
for each item weight, the corresponding profit is calculated by the formula presented in the first paragraph of this section.

\subsubsection{CSP pricing subproblem dataset}
\label{sec:csp_ukp_inst}

An often mentioned UKP application is to solve the pricing subproblems generated by the linear programming relaxation of the Set Covering Formulation (SCF) for the Bin Packing Problem (BPP) and Cutting Stock Problem (CSP) using the column generation approach~\citep[p. 455--459]{book_ukp_2004}\citep{gg-61}.
To analyze the performance of the algorithms in the context of this application, the authors have written a C++ program that uses the CPLEX Solver to solve the SCF and feed the pricing problems generated to a custom UKP solving algorithm.

The experiments included eight datasets of BPP/CSP instances. The datasets are: Falkenauer (160 instances), Scholl (1210 instances), Wäscher (17 instances), Schwerin (200 instances), Hard28 (28 instances), Randomly Generated Instances (3840 instances), Augmented Non IRUP and Augmented IRUP Instances (ANI\&AI, 500 instances), and Gschwind and Irnich instances (GI instances, 240).
These datasets amount to 6195 instances, all made available in~\cite{Delorme2018}. %\url{http://or.dei.unibo.it/library/bpplib}. 
The first seven datasets are described in~\cite{survey2014}, the last one (GI Instances) comes from~\cite{irnich}.
The code used to solve the SCF relaxation can be found in the first author's repository\footnote{The C++/CPLEX code used for solving SCF relaxation is available at \url{https://github.com/henriquebecker91/masters/tree/8367836344a2f615640757ffa49254758e99fe0a/codes/cpp}. The code can be compiled by executing \emph{make bin/cutstock} in the folder.
The dependencies are the Boost C++ library (see: \url{http://www.boost.org/}), and IBM ILOG CPLEX Studio 12.5 (see: \url{https://www.ibm.com/developerworks/community/blogs/jfp/entry/cplex_studio_in_ibm_academic_initiative?lang=en})}.

\subsubsection{Datasets deliberately ignored}

The \emph{uncorrelated} and \emph{weakly correlated} item distributions were commonly used in the literature~\citep{mtu1,mtu2,babayev,eduk}, but the authors decided to not include them in the experiments.
The literature has already questioned the suitability of \emph{uncorrelated} item distributions datasets for the analysis of the UKP~\citep{zhu_dominated, ukp_new_results}.
Uncorrelated instances often exhibit a vast amount of simple and multiple dominated items, and polynomial algorithms can reduce the number of items in such instances by orders of magnitude.
%In the author's experience, uncorrelated instances often take more time to load from disk than to solve.
%The solving times of uncorrelated instances are more dependent on the implementation of polynomial-time preprocessing than dependent on the quality of the solving algorithm.
%Consequently, the authors do not believe uncorrelated instances provide a good benchmark for UKP solving algorithms.

The \emph{weakly correlated} item distribution can be seen as a \emph{strongly correlated} item distribution with more dominated items.
The authors found redundant to present weakly correlated datasets in addition to the strongly correlated datasets, as preliminary results suggested that the time spent solving weakly correlated datasets was similar to the time spent solving strongly correlated datasets of smaller size.

\subsection{Computer setup}

All experiments were run using a computer with the following characteristics:
the CPU was an Intel\textsuperscript{\textregistered} Core\textsuperscript{TM} i5-4690 CPU @ 3.50GHz;
there were 8GiB RAM available (DIMM DDR3 Synchronous 1600 MHz) and three levels of cache (256KiB, 1MiB, and 6MiB, with the cores sharing only the last one).
The operating system used was GNU/Linux 4.7.0-1-ARCH x86\_64 (i.e., Arch Linux). 
Three of the four cores were isolated using the \emph{isolcpus} kernel flag (the non-isolated core was left to run the operating system). 
The \emph{taskset} utility was used to execute runs in one of the isolated cores.
All runs were executed in serial order, as the authors found that parallel executions effected the run times, even if each isolated core only hosted one run at each time~\citep[p.~87]{ukp_hb_mastersthesis}.

The OCaml code (PYAsUKP/EDUK/EDUK2) was compiled with ocamlopt and the flags suggested by the authors of the code for maximum performance (\emph{-unsafe -inline 2048}).
The Fortran code (original MTU1/MTU2) was compiled with gcc-fortran and \emph{-O3 -std=f2008} flags enabled.
The C++ code (all remaining implementations) was compiled with gcc and the \emph{-O3 -std=c++11} flags enabled.

\section{Results and Analyses}
\label{sec:exp_and_res}

The experiments are split into five subsections.
Each section addresses one dataset and the results of running some selected algorithms over them.
To keep the results close to its discussion, each experiment section brings both the results and their immediate analysis.
Only implementations including the tiebreaker improvement were used in the experiments, thus, for simplicity, the Revisited Ordered Step-Off, Revisited Terminating Step-Off and Revisited GFDP will be referred as OSO, TSO and GFDP in this section.
%The big picture painted by all the experiments is discussed in \autoref{sec:discussion}.

\subsection{Results on the PYAsUKP dataset}
\label{sec:pya_exp}

The experiment presented in this section is an updated version of the experiment first presented in~\cite{sea2016}.
In this version GFDP is considered, UKP5 is replaced by TSO, and all runs were executed serially.
The same 4540 instance files were used.

\begin{figure}[!htbp]
\caption{Run times of TSO, GFDP and EDUK2 algorithms over the 4540 instances of the PYAsUKP dataset. There was no time limit. The instance classes acronyms stand for Postponed Periodicity (PP); Strongly Correlated (SC); Subset-Sum (SS); and Without Collective Dominance (WCD). 
EDUK was not included because EDUK2 supersedes EDUK. The ordered step-off run times were omitted because they were too similar to the terminating step-off run times. The GFDP run times over subset-sum instances are not displayed because in this case the two most efficient items have the same efficiency and, therefore, GFDP behaves as OSO. The mean labels inform the mean run times of each algorithm for the corresponding dataset, in seconds.}
\begin{center} 
\begin{knitrout}
\definecolor{shadecolor}{rgb}{0.969, 0.969, 0.969}\color{fgcolor}
\includegraphics[width=\maxwidth]{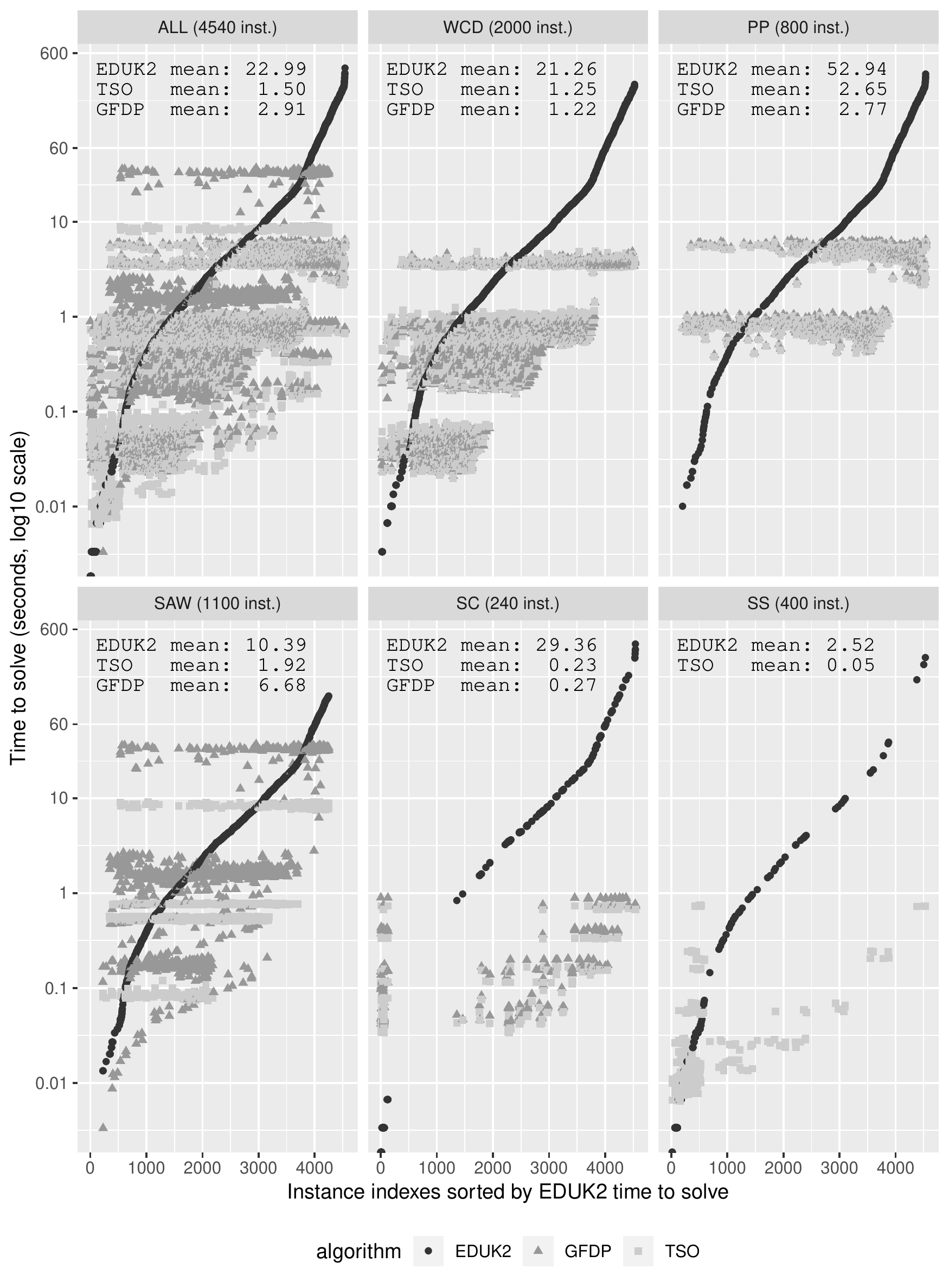} 

\end{knitrout}

\end{center}
%\legend{Source: the author.}
\label{fig:pya_fast}
\end{figure}

In \autoref{fig:pya_fast}, the instances (\(x\)-axis) are sorted by the time EDUK2 spent to solve them.
EDUK2 B\&B phase allows EDUK2 to solve some instances (distributed between all instance sizes) faster than TSO/GFDP.
This behavior shows that EDUK2 times for a specific instance cannot be predicted based on the number of items and distribution of the instance.
Nonetheless, when the B\&B phase has little effect, EDUK2 DP phase (basically EDUK) spend considerably more time than TSO/GFDP to solve the instance.

TSO and GFDP run times form plateaus in the figure.
For each class of instances, the plateaus aggregate runs over instances with a number of items of similar magnitude.
This behavior shows that the specific items that constitute an instance affect TSO and GFDP less than the number of items and distribution, both which are good predictors of TSO and GFDP run times.

The run times of TSO and GFDP are similar, except for the SAW instances, in which GFDP performed considerably worse than TSO.
GFDP DP phase does not generate solutions including the best item.
The use of the best item allows the generation of more efficient solutions, which dominate less efficient solutions, reducing the total number of solutions generated and, consequently, the computational effort spent.
The authors believe that the exclusion of the best item from the DP phase is the reason that GFDP presented run times higher than TSO over SAW instances.

When EDUK2 solves an instance in less time than TSO, the difference is often less than a second, and up to ten seconds.
When EDUK2 solves an instance in more time than TSO, the difference is often more than five seconds and up to six minutes.
Such behavior harms EDUK2 mean run time in comparison to simpler DP methods.

\subsubsection{MTU1 and MTU2 (C++ and Fortran)}
\label{sec:mtu_exp}

This section compares the performance of the C++ and Fortran implementations of MTU1 and MTU2 algorithms.
These four implementations were executed over the reduced PYAsUKP benchmark.

\begin{figure}[!htbp]
\caption{Run times of MTU1 and MTU2 (C++ and Fortran) over the 454 instances of the reduced PYAsUKP dataset. The time limit was set to 30 minutes. Runs terminated by timeout are displayed as taking exactly the time limit.}
\begin{center}
\begin{knitrout}
\definecolor{shadecolor}{rgb}{0.969, 0.969, 0.969}\color{fgcolor}
\includegraphics[width=\maxwidth]{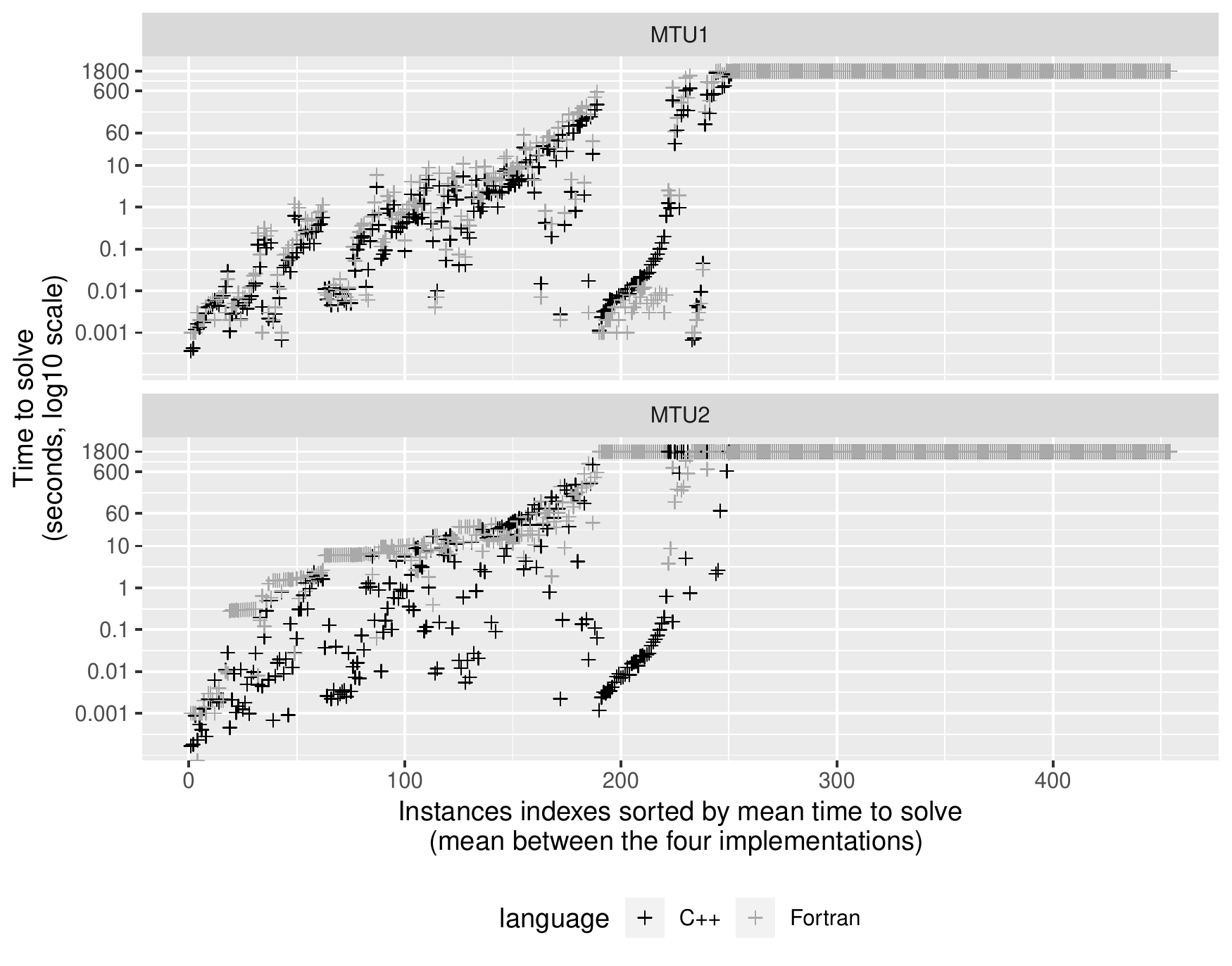} 

\end{knitrout}
\end{center}
%\legend{Source: the author.}
\label{fig:mtu}
\end{figure}

In \autoref{fig:mtu}, both MTU1 implementations show run times in the same order of magnitude.
However, considering only the instances that both MTU1 implementations solved before the timeout, the mean run time of Fortran MTU1 was 59 seconds and the mean run time of C++ MTU1 was 30 seconds.
The analysis of the individual run times shows that for many instances Fortran MTU1 spent about the double of the time spent by C++ MTU1 to solve the same instance.

For many instances, the run times of the MTU2 implementations differed in more than one order of magnitude.
The authors believe that this divergence was caused by the difference of sorting algorithms and items ordering (C++ MTU2 order items by nondecreasing weight if they share the same efficiency).
The subset-sum instances were the ones that exhibited the largest difference.
The range \([190,221]\) of the \(x\)-axis of \autoref{fig:mtu} is composed of subset-sum instances.
C++ MTU2 solved all 40 subset-sum instances with a mean time of \(0.04\) seconds while Fortran MTU2 solved only eight instances with a mean time of \(155\) seconds.
Disregarding the subset-sum instances, the mean run time of Fortran MTU2 was 56 seconds and the mean run time of C++ MTU2 was 40.5 seconds.
Only the C++ implementations are used in the rest of the experiments.

%The subset-sum instances are always naturally sorted by efficiency, as all items share the same efficiency.
%The instances generated by PYAsUKP are also sorted by non-decreasing item weight, what makes them already sorted to C++ MTU2.
%The Fortran MTU2 does not seem to work well with this item ordering of the subset-sum instances.

\subsubsection{CPLEX and Gurobi}
\label{sec:cplex_gurobi_exp}

As in last section, this section uses the reduced PYAsUKP benchmark to gauge the relative performance of two alternatives, in this case: our Gurobi and CPLEX models.

\begin{figure}[!htp]
\caption{Run times of CPLEX and Gurobi over the 454 instances of the reduced PYAsUKP dataset. The time limit was set to 30 minutes. Runs terminated by timeout, memory exhaustion, or with wrong results are displayed as taking exactly the time limit.}
\begin{center}
\begin{knitrout}
\definecolor{shadecolor}{rgb}{0.969, 0.969, 0.969}\color{fgcolor}
\includegraphics[width=\maxwidth]{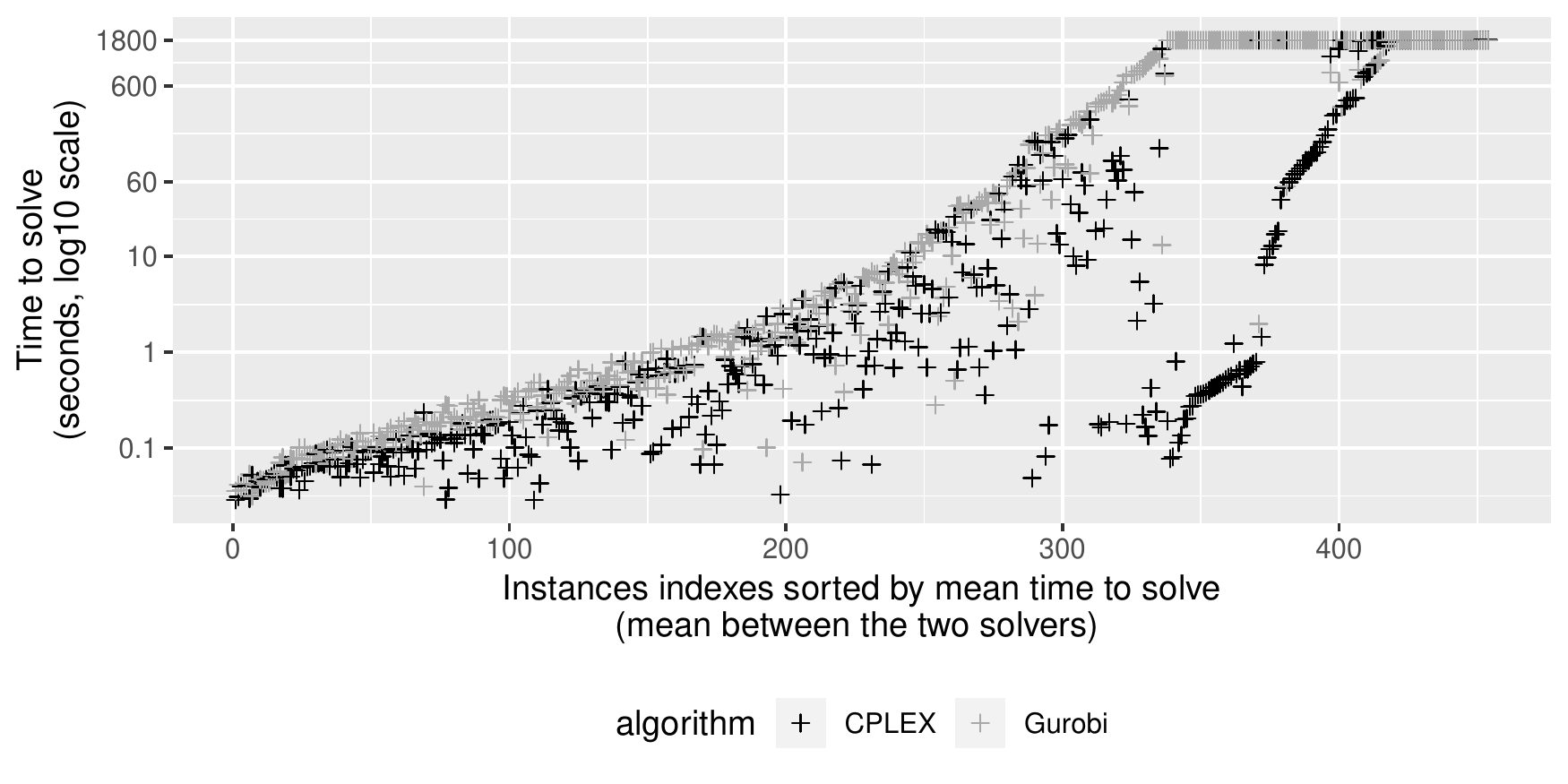} 

\end{knitrout}
\end{center}
%\legend{Source: the author.}
\label{fig:cplex_gurobi}
\end{figure}

Both Gurobi and CPLEX have an emphasis/focus setting.
The authors selected the emphases they deemed more adequate, and performed this experiment with all of them.
For each solver, only the results for the emphasis with the best performance are presented\footnote{
  The focus/emphasis parameters are GRB\_IntParam\_MIPFocus (Gurobi) and IloCplex::Param::Emphasis::MIP (CPLEX).
  The default of both solvers is balance between feasibility and optimality (code 0).
  Besides the default value (Gurobi best performance), the focus on optimality (code 2 on both solvers, CPLEX best performance) and the aggressive focus on optimality (code 3 on both solvers) were also tested.
  Other foci, as the emphasis in feasibility, were ignored as they did not seem relevant for an UKP model.
}.

Gurobi returned slightly wrong results for three instances (in which it warned about integrality tolerance violation) and exhausted the available memory (more than 7GiB) in another fifteen instances.
CPLEX did not present the same issues.
Performance-wise, \autoref{fig:cplex_gurobi} shows a clear advantage of CPLEX over Gurobi, not only with slightly shorter times for most instances, but there were also many cases in which Gurobi ended in timeout (or near it) and CPLEX finished in orders of magnitude less time.
Even only considering the finished and correct runs of both solvers, CPLEX mean time was 57 seconds while Gurobi mean time was 94 seconds.
None of the solvers can compete with TSO/GFDP/EDUK2 (as seen in \autoref{sec:pya_exp}), however their performance is superior to MTU1 and MTU2 (as seen in \autoref{sec:mtu_exp}).
Taking in account Gurobi performance and its memory and integrality issues, the authors chose to restrict the solvers presented to only CPLEX for the remaining experiments (the same emphasis was kept, unless said otherwise).

\subsection{Results on the Realistic Random Dataset}
\label{sec:rr_exp}

\begin{figure}[ht]
\caption{Run times of the algorithms over the 80 instances of the realistic random dataset. The time limit was set to 30 minutes. Runs terminated by timeout are displayed as taking exactly the time limit. Shape and color are used to distinguish between algorithms. The shape size is used to indicate the amount of overlapping points. The horizontal position of the points was adjusted for better visualization.}
\begin{center}
\begin{knitrout}
\definecolor{shadecolor}{rgb}{0.969, 0.969, 0.969}\color{fgcolor}
\includegraphics[width=\maxwidth]{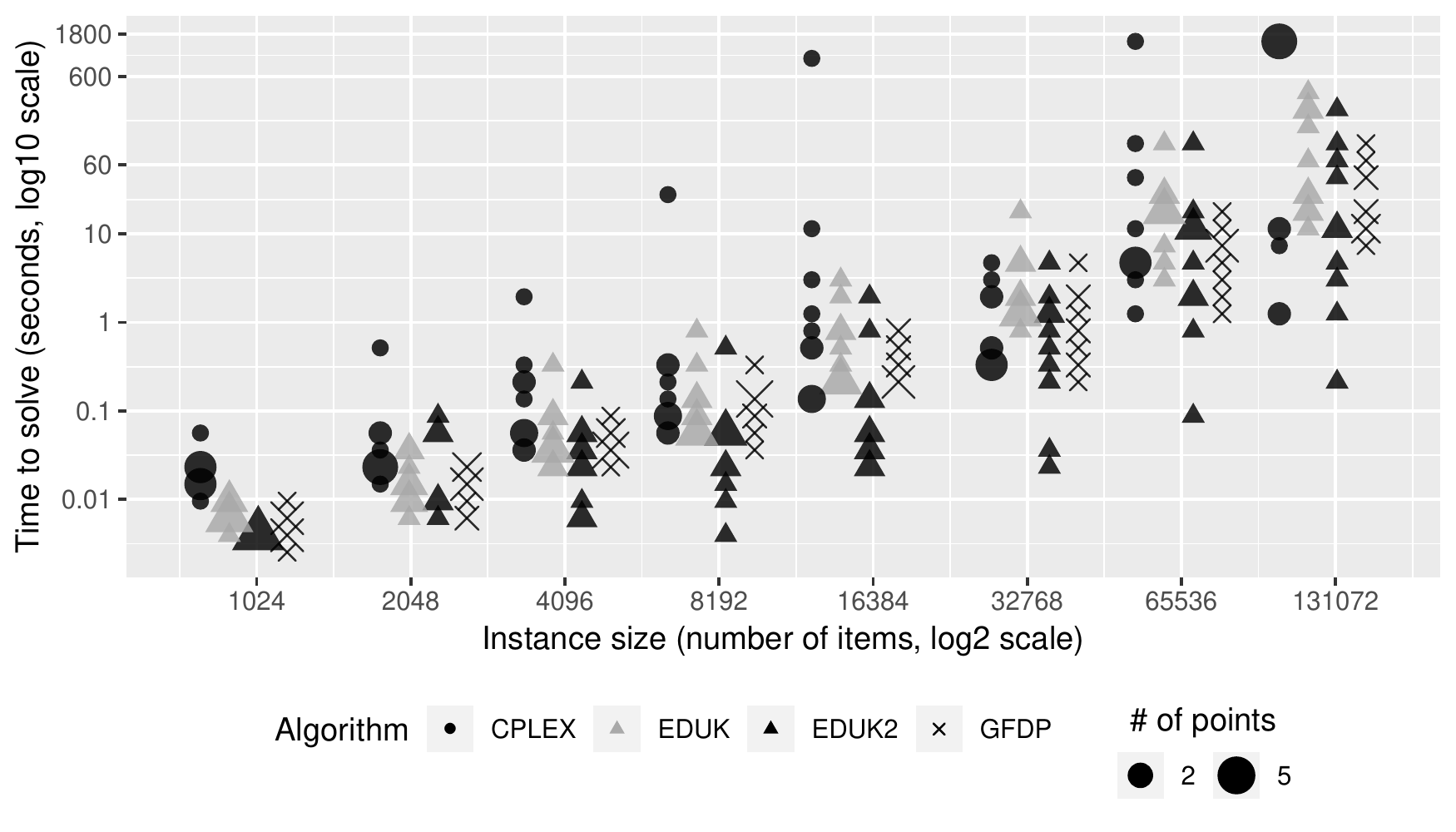} 

\end{knitrout}
\end{center}
%\legend{Source: the author.}
\label{fig:realistic_random}
\end{figure}

The results of the experiments over the realistic random dataset are summarized in \autoref{fig:realistic_random}.

The run times of both step-off algorithms and GFDP become almost identical as the size of the input grows, so the authors chose to present only the results of GFDP.
As the instance size grow, EDUK run times get a bit worse than GFDP.
EDUK2 has many run times similar to EDUK but, for some instances in each instance size, its B\&B phase solves the instance or helps to reduce the instance size considerably.
The B\&B methods (CPLEX, MTU1 and MTU) had similar a similar spreading behaviour, and the authors chose CPLEX as their best representative.
MTU1 and MTU2 had the best run times for smaller instances but, as the instance size and \(w_{min}\) grew, they had more timeouts than finished runs by \(n = 32768\) and only one finished run for the largest instance size.
%the shortest run times for the smallest instances but, as the instance size and \, also grow the number of instances in which MTU1 and MTU2 spend orders of magnitude more time to solve than the other algorithms.
%The pure B\&B algorithms (MTU1 and MTU2) show more variation than any other algorithms.

Despite EDUK2 solving some instances orders of magnitude faster than the other algorithms (especially in the larger instance sizes), the mean run time of EDUK2 (8.51 seconds) is higher than TSO mean run time (5.36 seconds).
As already observed in \autoref{sec:pya_exp}, EDUK2 often presents a few run times that are one order of magnitude higher than the highest run time of TSO, what considerably increases its mean time.

\subsection{Results on the BREQ 128-16 Standard Benchmark}
\label{sec:breq_exp}

\begin{figure}[!htbp]
\caption{Run times of the algorithms over the 100 instances of the BREQ 128-16 Standard Benchmark. The time limit was set to 30 minutes. Runs terminated by timeout are displayed as taking exactly the time limit. Shape and color are used to distinguish between algorithms. The shape size is used to indicate the amount of overlapping points. The horizontal position of the points was adjusted for better visualization.}
\begin{center}
\begin{knitrout}
\definecolor{shadecolor}{rgb}{0.969, 0.969, 0.969}\color{fgcolor}
\includegraphics[width=\maxwidth]{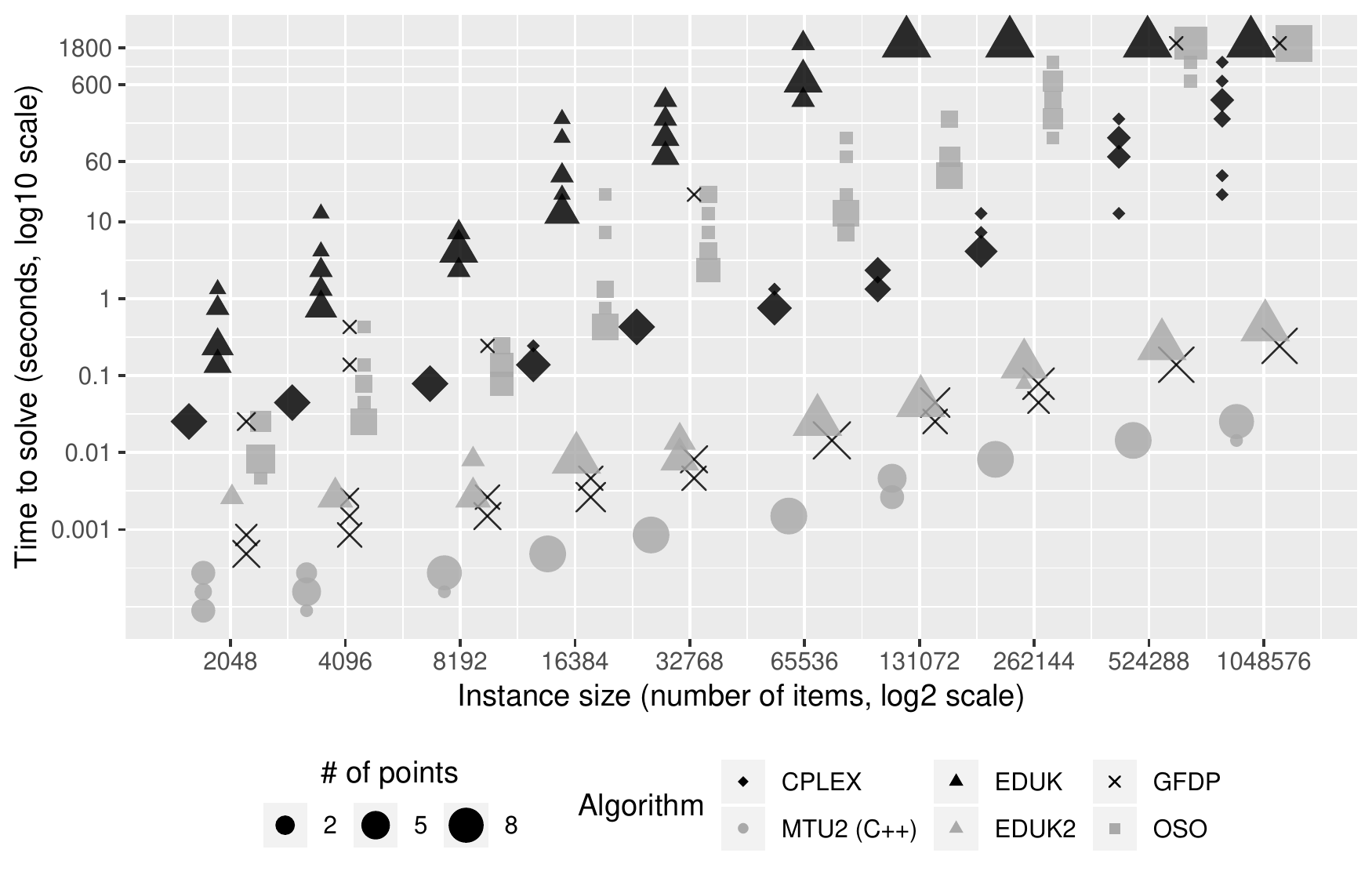} 

\end{knitrout}

\end{center}
%\legend{Source: the author.}
\label{fig:breq}
\end{figure}

The results for the BREQ 128-16 Standard Benchmark are summarized in \autoref{fig:breq}.
The run times outline two groups with distinct time growth: a steep-growth group (which hit the time limit) and a gradual-growth group (in which no run ends in timeout).
The steep-growth group is mainly composed of the DP algorithms: TSO, OSO, and EDUK.
The gradual-growth group is mainly composed of the B\&B and hybrid algorithms: CPLEX, MTU1, MTU2, and EDUK2.
MTU1 is not shown because its results are similar but dominated by MTU2, the analogue applies to TSO and OSO.
CPLEX is classified as gradual-growth, it is orders of magnitude worse than its fellow group members, but yet orders of magnitude better than the steady-growth group members, and has no run ending in timeout.
GFDP is the only algorithm with run times in both groups.

As EDUK2 is basically EDUK plus a B\&B phase and bounds checking, seems clear that it is the B\&B hybridization that allows for EDUK2 to be in the gradual-growth group (while EDUK that is pure DP is in the steep-growth group).
As the difference between MTU1 and MTU2 is the core strategy, which allows MTU2 to avoid sorting the entire item list, the time spent by these algorithm to sort the items is significant when executed over BREQ instances.
GFDP is an OSO variant which periodically computes bounds (similar to the ones used by the B\&B approach) to verify if the DP can be stopped and any remaining capacity be filled with copies of the best item.
If the bounds stop the DP early then GFDP run time falls in the gradual-growth group.
Otherwise, the run time is similar to OSO run time for the same instance.
CPLEX run times are explained by the fact that they are mostly spent solving the relaxation of the root node (i.e., struggling with the large number of variables), virtually no time is spent in the B\&B phase (often the B\&B tree has a single node).

The results confirm the hypothesis that instances generated with this distribution would be hard to be solved by DP algorithms and easy to be solved by B\&B algorithms.

\subsection{Results on the pricing subproblems from BPP/CSP}
\label{sec:csp_experiments}

The previous experiments included datasets of UKP instances and binaries that read and solved a single instance and then returned the solving time.
In this experiment, the instances are BPP/CSP instances.
Also, the run time for each BPP/CSP instance presented is the sum of all time spent solving the multiple pricing subproblems generated by the column generation approach applied over the linear programming relaxation of the set covering formulation.
In this experiment, the CPLEX version used was 12.5, and the emphasis is the default.

% The TSO, GREENDP, MTU2 (C++ or Fortran), and EDUK/EDUK2 (PYAsUKP) implementations were not used in this experiment.
Only CPLEX, OSO and MTU1 (C++) were used as pricing problem solvers in this experiment.
The values of \(n\) and \(c\) in the pricing problems are sufficiently small to exist little difference between GFDP/TSO compared to OSO, or MTU2 compared to MTU1, and therefore GFDP, TSO and MTU2 were removed from the comparison.
CPLEX was orders of magnitude slower than the other two and its results are presented only in the supplementary material to simplify the analysis.
The authors attribute the CPLEX poor performance to the same difficulties than MTU1 (arisal of `hard' problems) but also to a costly initialization of the solving process (each `easy' problem costed much more to CPLEX than to MTU1).
The authors did not succeed in integrating the PYAsUKP code (written in OCaml) with the C++/CPLEX code needed by this experiment.

%GREENDP is the same as OSO if the two most efficient items share the same efficiency, what often happens for at least one pricing subproblem of a CSP instance.
%In instances with a small number of items, MTU2 behaves almost as MTU1, the same applies to TSO and OSO.
%Since the instance capacity for these instances are small, GFDP has a similar performance than OSO.
%CPLEX was also used to solve the pricing problems, adding one more algorithm to the comparison.

In a pricing subproblem, the profit of the items is a real number.
Adapting MTU1 for using floating point profit values is not trivial.%, as the bound computation procedure is based on the assumption that both weight and profit values are integers.
The solution found was to multiply the items profit values by a multiplicative factor, round them down and treat them as integer profit values.
The multiplicative factor chosen was~\(2^{40}\) (approx. \(10^{12}\)).
In a pricing subproblem, the profit of the items can also be non-positive, which breaks the assumptions of some algorithms.
Items with non-positive profits are removed from the item list before passing it to the UKP solving algorithm.

\begin{figure}[!htbp]
\caption{Time spent solving the pricing subproblems from the 6195 CSP instances, with MTU1 and OSO (weight, integer). % three selected algorithms.
The time limit was set to 30 minutes (the time limit considered the total run time, not only the time spent solving pricing subproblems). If a run is terminated by timeout, the time spent solving pricing subproblems is displayed as exactly the time limit. The labels mean: n -- number of instances in the dataset; OSO/MTU1 n -- number of instances solved before timeout by the algorithm; OSO/MTU1 mean -- mean of the algorithm run times that did not end in timeout.}
\begin{center}

%<<knapsack_time1, fig.height = 2 >>=
%csp_csv$dataset <- 'All datasets'
%plot_csp(csp_csv) +
%theme(legend.position = 'none') +
%xlab('') +
%ylab('')
%@

\begin{knitrout}
\definecolor{shadecolor}{rgb}{0.969, 0.969, 0.969}\color{fgcolor}
\includegraphics[width=\maxwidth]{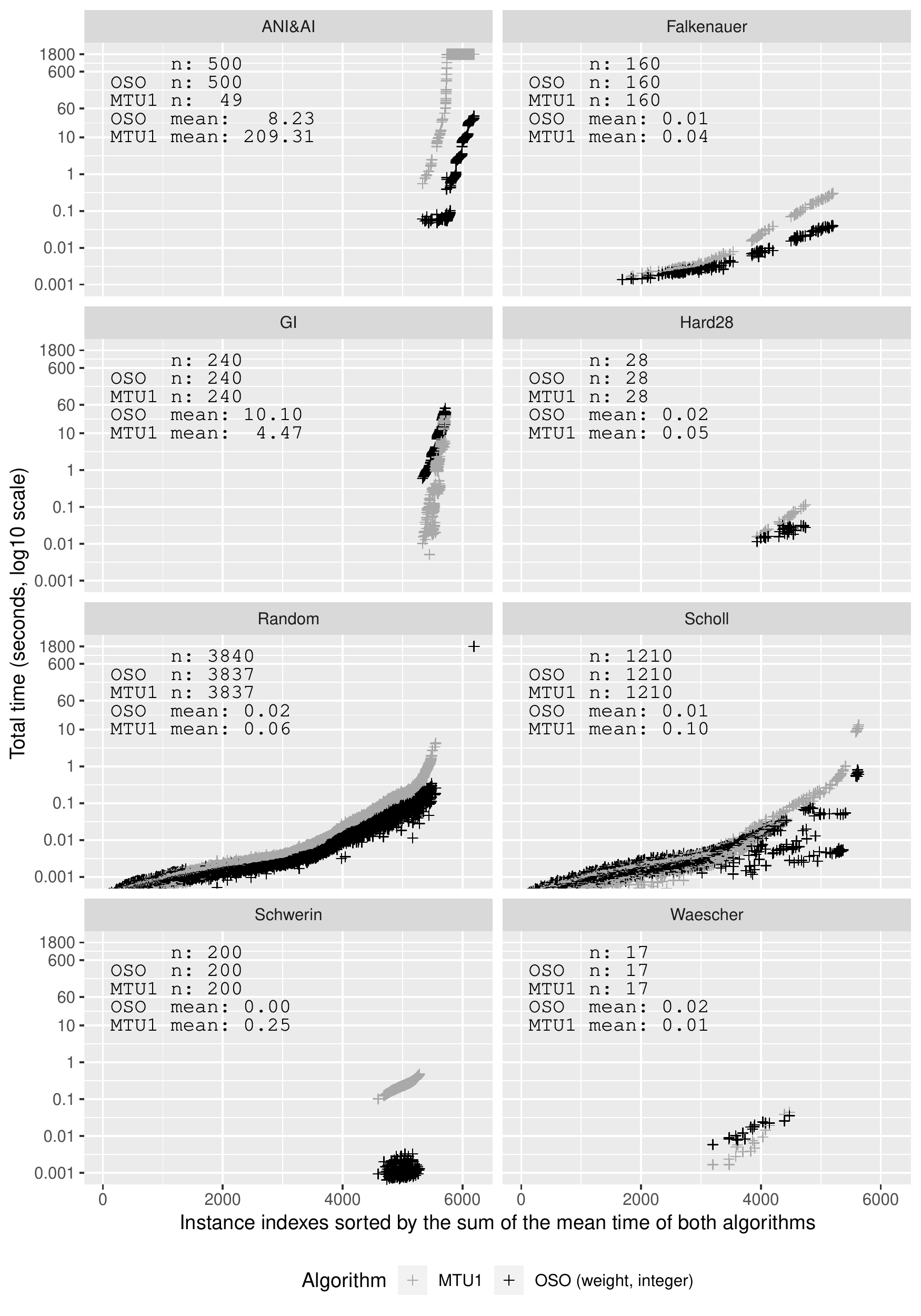} 

\end{knitrout}
\end{center}
%\legend{Source: the author.}
\label{fig:csp_knapsack_time}
\end{figure}

In \autoref{fig:csp_knapsack_time}, it can be seen that, except by two recent datasets (ANI\&AI and GI), the mean time spent solving pricing problems in an instance is below one second.
Such times are explained by two main factors: (1) for completeness, the authors included many classic but old datasets which nowadays are easy to solve; 
(2) these BPP/CSP datasets were meant to be hard to solve exactly, and solving the linear programming relaxation of the problem takes only a fraction of that time.
In the majority of these easy datasets, the highest times presented are from MTU1, while OSO presented the lowest mean time.

Many instances of the ANI\&AI dataset exceeded the timeout when MTU1 was used to solve the pricing subproblems.
To study this behavior, next the times of individual pricing problems of the instance 1002\_80000\_DI\_12 (ANI\&AI dataset) are analyzed.
The pricing problems generated by the same BPP/CSP instance always share the same \(n\), \(c\) and items weights. The only difference between the pricing problems is the item profit values\footnote{The specific code used removes items with nonpositive profit values before the beginning of the algorithms. Consequently, \(n\) do vary, but the results are the same.}.
The pricing problems generated when solving this specific instance have \(n = 911\) and \(c = 66432\).

\begin{figure}[!htbp]
\caption{Time spent solving each individual pricing problem generated by instance 1002\_80000\_DI\_12 (ANI\&AI dataset) with MTU1 and TSO. The MTU1 run was killed by timeout (30 minutes), TSO run finished normally.}
\begin{center}
\begin{knitrout}
\definecolor{shadecolor}{rgb}{0.969, 0.969, 0.969}\color{fgcolor}
\includegraphics[width=\maxwidth]{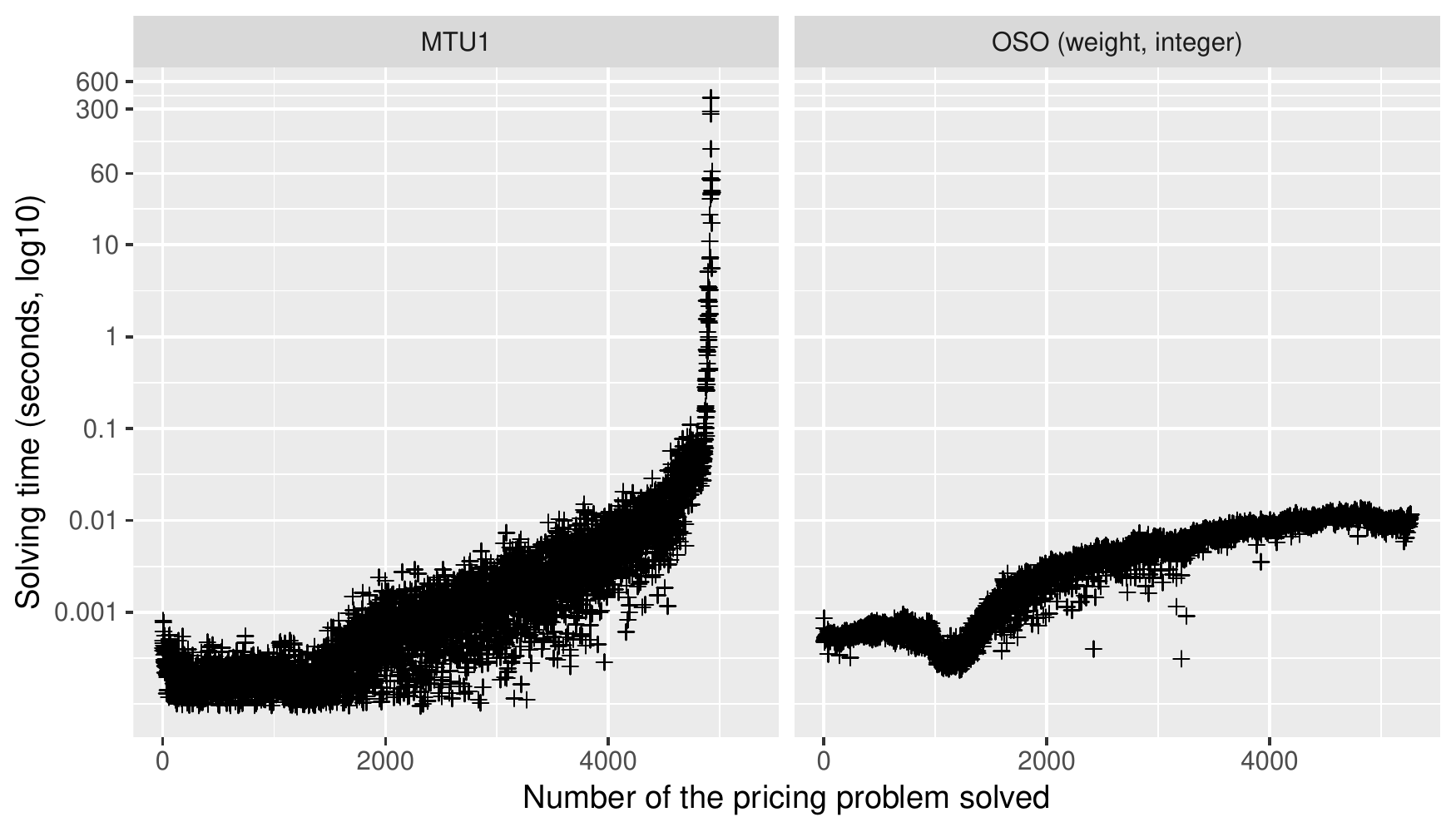} 

\end{knitrout}
\end{center}
%\legend{Source: the author.}
\label{fig:csp_inst}
\end{figure}

In \autoref{fig:csp_inst}, the times of pricing problems generated by this instance are presented.
It can be seen that the MTU1 run exceed the time limit because the time spent solving the last pricing problems increases exponentially.
The time taken by OSO to solve pricing problems also increases, but one or two orders of magnitude (not five or six).

\begin{figure}[!htbp]
\caption{Selected pricing problems generated while solving instance 1002\_80000\_DI\_12 (ANI\&AI dataset) linear programming relaxation with MTU1 as pricing problem solver. The titles of the facets follow the format \emph{number of the pricing problem (how many seconds MTU1 spent to solve it)}. For this figure the time limit set was \(90\) hours.}
\begin{center}
\begin{knitrout}
\definecolor{shadecolor}{rgb}{0.969, 0.969, 0.969}\color{fgcolor}
\includegraphics[width=\maxwidth]{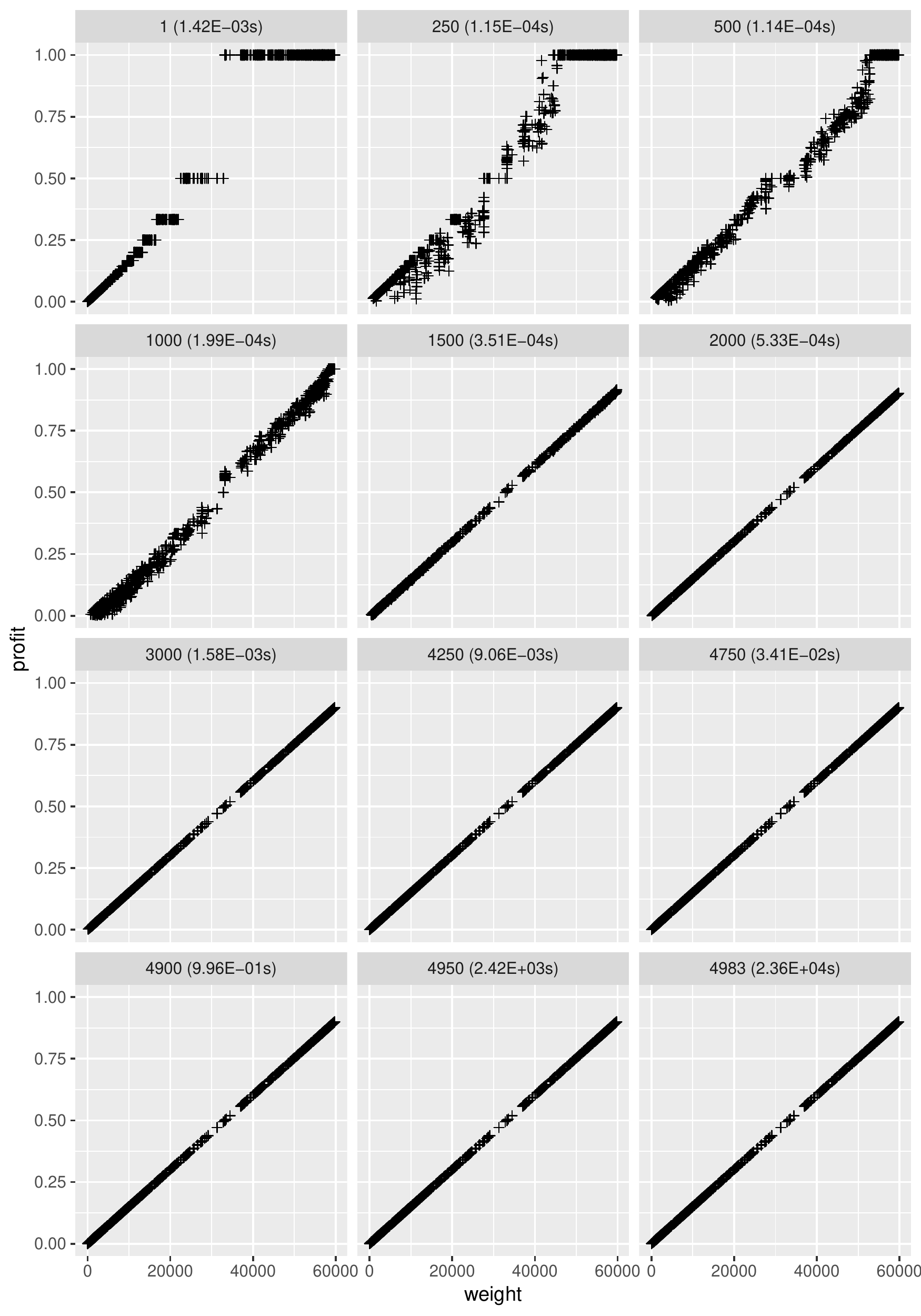} 

\end{knitrout}
\end{center}
\label{fig:twelve_pp}
\end{figure}

In \autoref{fig:twelve_pp}, it can be seen that as the profit values of the pricing problems change, their item distribution also change.
The initial distribution is consequence of the set of patterns used to initialize the column generation.
Given the bin size \(c\) and the \(n\) item sizes \(w_i\) (\(i=1,\dots,n\)), the initial set of patterns consists in one pattern for each item \(i\), with \(\floor{c/w_i}\) copies of that item.
Consequently, each item \(i\) with the same \(q = \floor{c/w_i}\) value has the same \(1/q\) profit value in the first iteration.
As patterns including items of different sizes are added, the profit of the items seems to approximate \(c/w_i\).
Instances with such distribution can be harder to solve by B\&B algorithms, as the similar efficiency weaken the capability of the bounds of reducing the solution space.
The authors also do not discard the possibility that using the multiplicative factor has reduced the quality of the MTU1 bounds (and its capability of finishing early) by cutting off some precision of the profit value.

To measure the impacts of the multiplicative factor, OSO variants using floating point profit values and using integer profit values were used in the experiments.
The weight of an item does not change from a pricing problem to the next, consequently, the items can be kept ordered by increasing weight with no effort.
The OSO algorithm sorts the items by nonincreasing efficiency but does not depend on this ordering to work.
To verify if the sorting cost would pay off, variants with sorting enabled and disabled were used.
Consequently, four versions of OSO were used in this experiment, for all combinations of profit type (the original floating point, or the converted integer), and sorting (sorting each new pricing problem by nonincreasing efficiency, or sorting the items by increasing weight a single time before the first pricing problem).

In the experiments with UKP instance datasets, all algorithms agreed on the value of the optimal solution, while sometimes returned different optimal solutions.
In this experiment, the difference in exactly which optimal solution is returned for a pricing subproblem (because of different tiebreaking or because of floating point inaccuracy) changes the next pricing problem and, consequently, the next optimal solution, in a feedback loop.
However, between MTU1 and the four OSO variants, for the same BPP/CSP instance, no solution of the linear programming relaxation (number of rolls needed) differed more than a \(2^{-18}\) fraction of a roll.
Solving pricing problems using the multiplicative factor to work over integer profits seems to be a viable approach. % a numerically stable approach.

%The time spent solving all pricing problems of a specific BPP/CSP instance the four OSO variants. %The distribution of the total time spent solving pricing problems for each BPP/CSP instance was similar between the four OSO variants.
The four OSO variants spent time in the same order of magnitude to solve all pricing problems of the same BPP/CSP instances.
The mean time of the four variants differed significantly: \(1.56\)s (efficiency, floating point), \(1.21\)s (weight, floating point), \(1.46\)s (efficiency, integer), \(1.06\)s (weight, integer).
Using integer profits show a small but consistent improvement in the times (the time used to convert the value is included), keeping the items in the natural increasing weight order shows a greater and also consistent improvement.
The mean time reduction observed in `weight' variants did not come from cutting the time spent sorting the items (less than 1\% of the total pricing time), but from how the DP phase of OSO behaved with a differently sorted list.
The times of OSO in all figures of this section are from the weight/integer variant.

\section{General Discussion and Conclusions}
\label{sec:discussion}

Except for the BREQ dataset, a revisited version of a DP algorithm from 1966 (Revisited Terminating Step-Off, R-TSO) had lower mean times than the only known implementation of the current state-of-the-art algorithm (EDUK2 /PYAsUKP).
Such results bring up two central questions the authors address: Why TSO was not considered in recent comparisons? How R-TSO outperformed the current state-of-the-art after five decades of study of the UKP?

The authors believe TSO was not considered in recent comparisons because: the algorithm was very similar to the naïve DP for the UKP but the authors did not emphasize it was orders of magnitude faster; B\&B algorithms performed better than TSO in uncorrelated and weakly correlated instances with one hundred items.
As far as the authors know, the proposal of MTU1 was the last time TSO was included in a comparison~\citep{mtu1}.
In the experiments presented in~\cite{mtu1}, TSO was about four times slower than MTU1 in the instances with~\(n = 25\); about two or three times slower in the instances with~\(n = 50\); and less than two times slower in instances with~\(n = 100\).
Such instances are now too small to consider, and the relative difference between TSO and MTU1 mean times was less than one order of magnitude apart and diminishing.
This trend hinted the possibility of the times taken by OSO/TSO and MTU1 converging (or even OSO/TSO taking less time than MTU1) for larger instances (e.g., OSO/TSO algorithm could have a costly initialization process but a better average-case complexity).
% Also, the trend observed could indicate that for instances with a greater \(n\) value, OSO/TSO algorithm would have lower times than MTU1, as their relative difference was diminishing. %(e.g., OSO/TSO algorithm could have a costly initialization process but a better average-case complexity).

\subsection{An algorithm is dominated by other in the context of a dataset}

The comparisons often found in the literature: (1) only compared a newly proposed algorithm to the algorithm that `won' the last comparison; (2) proposed new artificial datasets based on some definition of being `harder to solve'.
These two characteristics led to the following scenario: algorithm B dominates algorithm A in the context of dataset D1; algorithm C dominates algorithm B in the context of dataset D2; nothing guarantees that algorithm A does not dominate algorithm C in the context of dataset D2, as algorithm A was not included in the last comparison.
%The newly proposed dataset is often harder to solve because it has characteristics that made it harder to solve by some algorithm in the comparison (often the `loser').
%It is not considered if the characteristics that make it harder also negatively affect the algorithms excluded from the comparison, or if they are algorithm or approach-specific.

A concrete example of this behavior in the UKP literature follows: MTU2 was compared only to MTU1, and in the context of a new dataset of large \(n\) and rich in simple and multiple dominated items~\citep{mtu2}; EDUK and EDUK2 were compared only to MTU2, and in the context of new datasets with a smaller \(n\) but with less dominated items and higher \(w_{min}\)~\citep{eduk,pya}.
To be fair, the objective of these papers was to show how the newly proposed algorithm was not negatively impacted by some instance characteristics as the older algorithm was.
However, in such experiments, no old competitive algorithm of the same solving approach as the newly proposed algorithm (DP or B\&B) was included.

The purpose of the BREQ item distribution is to further illustrate how artificial datasets that favor one approach over another can be created.
The BREQ instances are hard to solve by DP algorithms and easy to solve by B\&B algorithms.
If they were the only instances considered, then MTU2 would be considered the best algorithm (as it can be seen in \autoref{tab:one_table}).
However, considering all remaining datasets, MTU2 often presents the worst time performance.
The BREQ instances do not suffer from the same richness of simple, multiple and collective dominated items that led uncorrelated instances to be criticized and abandoned.
However, threshold dominance is widespread in BREQ instances and, as far the authors know, no real-world instances follow the BREQ distribution.%, so the authors do not suggest their use in future experiments.
%The authors ask any future researchers to take in account the purpose for which BREQ instances were created (i.e., favor the B\&B approach) before including them in their experiments.

The effects of artificial instances in shaping what is considered the best algorithms for UKP is not limited to the instances proposed for the UKP.
\cite{irnich} created the GI instances to be harder to solve by their column generation implementation: 
``[...] generated new and harder CS instances. These are characterized by huge values for the capacity (to complicate the subproblems) and larger numbers of items with distinct lengths.''.
Their implementation used a DP algorithm to solve pricing problems.
The characteristics of these newly proposed BPP/CSP instances made the pricing problems harder to solve by a DP algorithm, but not necessarily by a B\&B algorithm, which is less affected by parameters like the knapsack capacity.
To evaluate which is the best UKP algorithm to solve pricing subproblems, the BPP/CSP instances had also to be representative of real-world instances; otherwise, another layer of bias is laid.
The ANI\&AI instances, which MTU1 had difficulties to solve the pricing problems, are also instances created with the purpose to be hard to solve~\citep{survey2014}.
The definition of hard to solve is different between the GI and ANI\&AI instances, as in ANI\&AI instances the objective is to make B\&B algorithms for the BPP/CSP struggle to prove the optimality of a solution for a BPP/CSP instance.
The way such characteristic makes the pricing problems from ANI\&AI instances harder to solve by MTU1 is not so clear as in the case of GI instances and DP algorithms.

\begin{table}[ht]
\def\arraystretch{1.1}

\begin{adjustbox}{max width=\textwidth, center}
\begin{tabular}{@{\extracolsep{4pt}}lrrrrrrrr@{}}

& \multicolumn{2}{c}{PYAsUKP\textsuperscript{a}} & \multicolumn{2}{c}{RR} & \multicolumn{2}{c}{BREQ} & \multicolumn{2}{c}{CSP\textsuperscript{b}}\\
\cline{2-3}\cline{4-5}\cline{6-7}\cline{8-9}
Method & \multicolumn{1}{c}{fin} & \multicolumn{1}{c}{avg} & \multicolumn{1}{c}{fin} & \multicolumn{1}{c}{avg} & \multicolumn{1}{c}{fin} & \multicolumn{1}{c}{avg} & \multicolumn{1}{c}{fin} & \multicolumn{1}{c}{avg}\\
%& fin & time & fin & time & fin & time & fin & time\\
\cline{1-1}\cline{2-3}\cline{4-5}\cline{6-7}\cline{8-9}
 R-TSO & 4540 & 1.50 & 80 & 5.36 & 83 & 111.64 & 6192 & 1.45 \\ 
  R-OSO & 4540 & 1.52 & 80 & 5.42 & 83 & 111.57 & 6192 & 1.07 \\ 
  R-GFDP & 4140 & 2.91 & 80 & 5.32 & 98 & 0.33 & -- & -- \\ 
  EDUK2 & 4540 & 22.99 & 80 & 8.51 & 100 & 0.09 & -- & -- \\ 
  EDUK & -- & -- & 80 & 18.97 & 59 & 164.74 & -- & -- \\ 
  MTU2 & 231 & 38.11 & 55 & 11.61 & 100 & 0.01 & -- & -- \\ 
  MTU1 & 249 & 54.11 & 58 & 20.82 & 100 & 0.02 & 5741 & 2.04 \\ 
  CPLEX & 411 & 56.70 & 75 & 34.03 & 100 & 51.64 & 5606 & 33.38 \\ 
  Gurobi & 349 & 94.44 & -- & -- & -- & -- & -- & -- \\ 
  
\hline
\end{tabular}
\end{adjustbox}
%\caption{Time (in seconds) for the PYAsUKP 4540 Instances (see Section \ref{sec:pya_exp}). Columns \textbf{n} and \(w_{min}\) values must be multiplied by \(10^3\) to obtain their true value. Let \(T\) be the set of run times reported by the R-TSO, GFDP or EDUK2 for the instance dataset described by a row. The meaning of the columns \textbf{avg}, \textbf{sd} and \textbf{max}, is, respectively, the arithmetic mean of \(T\), the standard deviation of \(T\), the maximum value of \(T\). For the subset-sum instances, there are ten instances for each possible combination of: \(w_{min} \in \{10^3, 5\times10^3, 10^4, 5\times10^4, 10^5\}\); \(w_{max} \in \{5\times10^5, 10^6\}\) and \(n \in \{10^3, 2\times10^3, 5\times10^3, 10^4\}\), totaling 400 instances.}
\caption{
The number of \textbf{fin}ished runs of a method (row) over a dataset (column), and the \textbf{av}era\textbf{g}e time (in seconds) of the finished runs (dashes mean the respective method was not used to solve the respective dataset).
%The runs not finished are mostly timeouts (all runs had a half hour time limit), but some (as the Gurobi ones) include memory exhaustion and wrong results.
%The dashes mean the respective method was not executed over the respective dataset.
\textsuperscript{a} The MTU1, MTU2, CPLEX, and Gurobi were executed only over the 454 instances of the reduced PYAsUKP dataset (not all 4540 instances as the remaining methods).
\textsuperscript{b} The times reported are for the following variants: R-TSO -- integer profit, sort by efficiency; R-OSO -- integer profit, sort by weight; CPLEX -- custom code reusing the model and only changing the profits (coefficients of the objective function) between the iterations. %In the CSP dataset, the 6192 instances are CSP instances, the run is considered finished if the CSP was solved to proven optimality before the time limit, and the time is the mean time spent by instance solving all their pricing problems with the respective method. Also in the CSP column: the TSO results come from the variant that converts profit to integers and sort items by efficiency; the OSO variant also converts profit to integers but sort by weight; the CPLEX results come from a custom code that reuses the same model and only changes the profits at each iteration. 
}
\label{tab:one_table}
\end{table}

\subsection{Comments on R-TSO and EDUK2 times gap}

The authors believe many factors allowed R-TSO to outperform EDUK2 PYAsUKP implementation.
Some of them are: the weak solution dominance implicitly applied by (R-)TSO seems to be as effective as the explicit simple, multiple, collective and threshold dominance applied by EDUK2; R-TSO has better space locality; R-TSO solution backtrack trades memory for time (while EDUK does the opposite).

%The authors did not try to directly compare the effectiveness of EDUK2 and R-TSO dominance approach, as such would need changes in EDUK2 code to allow gathering extra data.
%However, as R-TSO finishes early if only the best item could be added to the solution, this means that TSO can discard every other item by applying solution dominance, as EDUK2 does.

EDUK2 PYAsUKP implementation uses lazy lists to store solutions, while R-TSO uses an array.
A strongly correlated  instance (\(\alpha = -5\), \(n = 10000\), \(w_{min} = 110000\), \(c = 9008057\)) of the PYAsUKP dataset was the one EDUK2 spent more time to solve (416 seconds, R-TSO spent about one second to solve the same instance).
By the use of the \emph{perf} profiler, it was possible to verify that EDUK2 PYAsUKP implementation executed about \(1122\) instructions per cache miss, while R-TSO executed about \(288653\) instructions per cache miss.
The performance gain for using an array-based implementation was also observed in~\cite[p.~19]{irnich}, which tried to follow the approach suggested by EDUK in their pricing problem solver: ``In contrast, for UKP we found that a straightforward array-based implementation of the DP approach is faster than the list-based approach. We suspect that on a modern CPU, the smaller state graph of UKP can be accessed much faster (due to caching techniques) so that the solution of the UKP subproblems as they occur in the BP benchmark instances is possible in almost no (measurable) time.''.

%In~\cite{eduk}, it is said that EDUK retrieves all optimal solutions for an UKP instance.
%To do the same, R-TSO could not make use of the tiebreaking optimization mentioned in \autoref{sec:oso_and_sol_dom}, what would affect its performance considerably in strongly correlated and subset-sum datasets.
%Also, the application of the solution dominance would need to be less strict as, for example, for two items \(j\) and \(i\) that respect \(w_i < w_j\) and \(p_i = p_j\), \(j\) can yet be present in an optimal solution.

\subsection{Conclusions}

In conclusion,
(1) the choice of artificial instance datasets had an important role defining which algorithms were considered the best by the literature; 
(2) the simple, multiple, collective, and threshold dominance relations can be generalized to solution dominance, and the application of a weak version of it shows similar efficiency;
(3) there is evidence that the items distribution of pricing problems can converge to a strongly correlated distribution (which can take exponential time to solve by B\&B);
(4) there is evidence that converting the profit of pricing problems to large integers do not cause significant loss to the master problem objetive value;
(5) the development of new and tighter periodicity bounds is of little use to the improvement of the state-of-the-art algorithms for UKP;
(6) CPLEX has better performance than B\&B algorithms for UKP in instances with few items but `hard' to solve by B\&B approach, but it is worse when the instances are large but `easy' (many variables) or too many small instances (costly initialization).

%While the following quote was written in the context of the 0-1 KP, the author found it relevant to complement what was just said: ``Dynamic programming is one of our best approaches for solving difficult (KP), since this is the only solution method which gives us a worst-case guarantee on the running time, independently on whether the upper bounding tests will work well.''~\cite[p.~13]{where_hard}.

\subsection{Future work}
\label{sec:future_works}

Many questions raised during the development of this paper ended up unanswered:% The authors selected they found more interesting in the list below.

\begin{itemize}
\item How similar are the datasets of the UKP and the BPP/CSP presented in the literature to the ones existent in the real world? Do the instances found in the real-world favor some approaches over others?
\item If a B\&B phase was added to the terminating step-off (as in EDUK2), and a C++ and array-based implementation of EDUK2 was written, would they have a similar performance?
%\item What would be the practical performance of an implementation of the algorithm described in~\cite{babayev}?
%\item How do the traits of the optimal solution for the pricing subproblem affect the master problem? Does always returning an optimal solution with minimal weight has a negative effect? What about adding all patterns that improve the master problem solution, and not only the best pattern (i.e., an optimal solution)? Or, instead, adding all optimal solutions instead of only one? %\item Are the profit values (and, consequently, the items distributions) of the pricing subproblems uniform between similar BPP/CSP instances, or the same BPP/CSP instance, or in both cases? Is it possible that they converge to a specific distribution at each iteration of the column generation?
\item The difference in performance of MTU1/MTU2 and CPLEX/Gurobi over the PYAsUKP dataset comes from the fact MTU1/MTU2 are depth-first? What would be the performance of a best-bound B\&B made for the UKP?
\end{itemize}

\subsection{Acknowledgements}

The authors are thankful to the CAPES (Coordenação de Aperfeiçoamento de Pessoal de Nível Superior) for the PROEX 0487 Bolsa País 1653154 and FAPERGS (Fundação de Amparo á pesquisa do Estado do Rio Grande do Sul) for the Edital 02/2017 - PqG 27720.414.17857.23062017. The authors are also thankful to Vincent Poirriez for his help with PYAsUKP.

\section{References}
\bibliography{biblio}

\begin{thebibliography}{30}
\expandafter\ifx\csname natexlab\endcsname\relax\def\natexlab#1{#1}\fi
\providecommand{\url}[1]{\texttt{#1}}
\providecommand{\href}[2]{#2}
\providecommand{\path}[1]{#1}
\providecommand{\DOIprefix}{doi:}
\providecommand{\ArXivprefix}{arXiv:}
\providecommand{\URLprefix}{URL: }
\providecommand{\Pubmedprefix}{pmid:}
\providecommand{\doi}[1]{\href{http://dx.doi.org/#1}{\path{#1}}}
\providecommand{\Pubmed}[1]{\href{pmid:#1}{\path{#1}}}
\providecommand{\bibinfo}[2]{#2}
\ifx\xfnm\relax \def\xfnm[#1]{\unskip,\space#1}\fi
%Type = Article
\bibitem[{Andonov et~al.(2000)Andonov, Poirriez \& Rajopadhye}]{eduk}
\bibinfo{author}{Andonov, R.}, \bibinfo{author}{Poirriez, V.}, \&
  \bibinfo{author}{Rajopadhye, S.} (\bibinfo{year}{2000}).
\newblock \bibinfo{title}{Unbounded knapsack problem: Dynamic programming
  revisited}.
\newblock {\it \bibinfo{journal}{European Journal of Operational Research}\/},
  {\it \bibinfo{volume}{123}\/}, \bibinfo{pages}{394--407}.
  \DOIprefix\doi{10.1016/S0377-2217(99)00265-9}.
%Type = Inproceedings
\bibitem[{Andonov \& Rajopadhye(1994)}]{algo_tech_cut}
\bibinfo{author}{Andonov, R.}, \& \bibinfo{author}{Rajopadhye, S.}
  (\bibinfo{year}{1994}).
\newblock \bibinfo{title}{A sparse knapsack algo-tech-cuit and its synthesis}.
\newblock In {\it \bibinfo{booktitle}{International Conference on Application
  Specific Array Processors, 1994.}\/} (pp. \bibinfo{pages}{302--313}).
\newblock \bibinfo{organization}{IEEE}.
\newblock \DOIprefix\doi{10.1109/ASAP.1994.331794}.
%Type = Article
\bibitem[{Babayev et~al.(1997)Babayev, Glover \& Ryan}]{babayev}
\bibinfo{author}{Babayev, D.~A.}, \bibinfo{author}{Glover, F.}, \&
  \bibinfo{author}{Ryan, J.} (\bibinfo{year}{1997}).
\newblock \bibinfo{title}{A new knapsack solution approach by integer
  equivalent aggregation and consistency determination}.
\newblock {\it \bibinfo{journal}{INFORMS Journal on Computing}\/},  {\it
  \bibinfo{volume}{9}\/}, \bibinfo{pages}{43--50}.
  \DOIprefix\doi{10.1287/ijoc.9.1.43}.
%Type = Masterthesis
\bibitem[{Becker(2017)}]{ukp_hb_mastersthesis}
\bibinfo{author}{Becker, H.} (\bibinfo{year}{2017}).
\newblock {\it \bibinfo{title}{The Unbounded Knapsack Problem: a critical
  review}\/}.
\newblock Master's thesis Federal University of Rio Grande do Sul.
\newblock \URLprefix \url{http://hdl.handle.net/10183/163413}.
%Type = Inproceedings
\bibitem[{Becker \& Buriol(2016)}]{sea2016}
\bibinfo{author}{Becker, H.}, \& \bibinfo{author}{Buriol, L.~S.}
  (\bibinfo{year}{2016}).
\newblock \bibinfo{title}{{UKP5}: A new algorithm for the unbounded knapsack
  problem}.
\newblock In {\it \bibinfo{booktitle}{International Symposium on Experimental
  Algorithms}\/} (pp. \bibinfo{pages}{50--62}).
\newblock \bibinfo{organization}{Springer}.
\newblock \DOIprefix\doi{10.1007/978-3-319-38851-9_4}.
%Type = Article
\bibitem[{Cabot(1970)}]{cabot}
\bibinfo{author}{Cabot, A.~V.} (\bibinfo{year}{1970}).
\newblock \bibinfo{title}{An enumeration algorithm for knapsack problems}.
\newblock {\it \bibinfo{journal}{Operations Research}\/},  {\it
  \bibinfo{volume}{18}\/}, \bibinfo{pages}{306--311}.
  \DOIprefix\doi{10.1287/opre.18.2.306}.
%Type = Techreport
\bibitem[{Delorme \& Iori(2017)}]{eq_lb_delorme}
\bibinfo{author}{Delorme, M.}, \& \bibinfo{author}{Iori, M.}
  (\bibinfo{year}{2017}).
\newblock {\it \bibinfo{title}{Enhanced Pseudo-Polynomial Formulations for Bin
  Packing and Cutting Stock Problems}\/}.
\newblock \bibinfo{type}{Technical Report} Optimization Online.
\newblock \URLprefix
  \url{http://www.optimization-online.org/DB_HTML/2017/10/6270.html}.
%Type = Article
\bibitem[{Delorme et~al.(2016)Delorme, Iori \& Martello}]{survey2014}
\bibinfo{author}{Delorme, M.}, \bibinfo{author}{Iori, M.}, \&
  \bibinfo{author}{Martello, S.} (\bibinfo{year}{2016}).
\newblock \bibinfo{title}{Bin packing and cutting stock problems: Mathematical
  models and exact algorithms}.
\newblock {\it \bibinfo{journal}{European Journal of Operational Research}\/},
  {\it \bibinfo{volume}{255}\/}, \bibinfo{pages}{1 -- 20}. \URLprefix
  \url{http://www.sciencedirect.com/science/article/pii/S0377221716302491}.
  \DOIprefix\doi{https://doi.org/10.1016/j.ejor.2016.04.030}.
%Type = Article
\bibitem[{Delorme et~al.(2018)Delorme, Iori \& Martello}]{Delorme2018}
\bibinfo{author}{Delorme, M.}, \bibinfo{author}{Iori, M.}, \&
  \bibinfo{author}{Martello, S.} (\bibinfo{year}{2018}).
\newblock \bibinfo{title}{Bpplib: a library for bin packing and cutting stock
  problems}.
\newblock {\it \bibinfo{journal}{Optimization Letters}\/},  {\it
  \bibinfo{volume}{12}\/}, \bibinfo{pages}{235--250}. \URLprefix
  \url{https://doi.org/10.1007/s11590-017-1192-z}.
  \DOIprefix\doi{10.1007/s11590-017-1192-z}.
%Type = Article
\bibitem[{Fischetti \& Martello(1988)}]{partial_sort_martello}
\bibinfo{author}{Fischetti, M.}, \& \bibinfo{author}{Martello, S.}
  (\bibinfo{year}{1988}).
\newblock \bibinfo{title}{A hybrid algorithm for finding the kth smallest of n
  elements in o(n) time}.
\newblock {\it \bibinfo{journal}{Annals of Operations Research}\/},  {\it
  \bibinfo{volume}{13}\/}, \bibinfo{pages}{399--419}.
  \DOIprefix\doi{10.1007/BF02288326}.
%Type = Book
\bibitem[{Garfinkel \& Nemhauser(1972)}]{garfinkel}
\bibinfo{author}{Garfinkel, R.~S.}, \& \bibinfo{author}{Nemhauser, G.~L.}
  (\bibinfo{year}{1972}).
\newblock {\it \bibinfo{title}{Integer programming}\/}
  volume~\bibinfo{volume}{4}.
\newblock \bibinfo{publisher}{Wiley New York}.
%Type = Article
\bibitem[{Gilmore \& Gomory(1961)}]{gg-61}
\bibinfo{author}{Gilmore, P.~C.}, \& \bibinfo{author}{Gomory, R.~E.}
  (\bibinfo{year}{1961}).
\newblock \bibinfo{title}{A linear programming approach to the cutting-stock
  problem}.
\newblock {\it \bibinfo{journal}{Operations research}\/},  {\it
  \bibinfo{volume}{9}\/}, \bibinfo{pages}{849--859}.
  \DOIprefix\doi{10.1287/opre.9.6.849}.
%Type = Article
\bibitem[{Gilmore \& Gomory(1963)}]{gg-63}
\bibinfo{author}{Gilmore, P.~C.}, \& \bibinfo{author}{Gomory, R.~E.}
  (\bibinfo{year}{1963}).
\newblock \bibinfo{title}{A linear programming approach to the cutting stock
  problem -- part {II}}.
\newblock {\it \bibinfo{journal}{Operations research}\/},  {\it
  \bibinfo{volume}{11}\/}, \bibinfo{pages}{863--888}.
  \DOIprefix\doi{10.1287/opre.11.6.863}.
%Type = Article
\bibitem[{Gilmore \& Gomory(1966)}]{gg-66}
\bibinfo{author}{Gilmore, P.~C.}, \& \bibinfo{author}{Gomory, R.~E.}
  (\bibinfo{year}{1966}).
\newblock \bibinfo{title}{The theory and computation of knapsack functions}.
\newblock {\it \bibinfo{journal}{Operations Research}\/},  {\it
  \bibinfo{volume}{14}\/}, \bibinfo{pages}{1045--1074}.
  \DOIprefix\doi{10.1287/opre.14.6.1045}.
%Type = Article
\bibitem[{Greenberg(1986)}]{on_equivalent_greenberg}
\bibinfo{author}{Greenberg, H.} (\bibinfo{year}{1986}).
\newblock \bibinfo{title}{On equivalent knapsack problems}.
\newblock {\it \bibinfo{journal}{Discrete applied mathematics}\/},  {\it
  \bibinfo{volume}{14}\/}, \bibinfo{pages}{263--268}.
  \DOIprefix\doi{10.1016/0166-218X(86)90030-2}.
%Type = Article
\bibitem[{Greenberg \& Feldman(1980)}]{green_improv}
\bibinfo{author}{Greenberg, H.}, \& \bibinfo{author}{Feldman, I.}
  (\bibinfo{year}{1980}).
\newblock \bibinfo{title}{A better step-off algorithm for the knapsack
  problem}.
\newblock {\it \bibinfo{journal}{Discrete Applied Mathematics}\/},  {\it
  \bibinfo{volume}{2}\/}, \bibinfo{pages}{21--25}.
  \DOIprefix\doi{10.1016/0166-218X(80)90051-7}.
%Type = Article
\bibitem[{Gschwind \& Irnich(2016)}]{irnich}
\bibinfo{author}{Gschwind, T.}, \& \bibinfo{author}{Irnich, S.}
  (\bibinfo{year}{2016}).
\newblock \bibinfo{title}{Dual inequalities for stabilized column generation
  revisited}.
\newblock {\it \bibinfo{journal}{INFORMS Journal on Computing}\/},  {\it
  \bibinfo{volume}{28}\/}, \bibinfo{pages}{175--194}.
  \DOIprefix\doi{10.1287/ijoc.2015.0670}.
%Type = Misc
\bibitem[{Gurobi~Optimization(2018)}]{gurobi}
\bibinfo{author}{Gurobi~Optimization, L.} (\bibinfo{year}{2018}).
\newblock \bibinfo{title}{Gurobi optimizer reference manual}.
\newblock \URLprefix \url{http://www.gurobi.com}.
%Type = Techreport
\bibitem[{Hu(1969)}]{tchu}
\bibinfo{author}{Hu, T.~C.} (\bibinfo{year}{1969}).
\newblock {\it \bibinfo{title}{Integer programming and network flows}\/}.
\newblock \bibinfo{type}{Technical Report} DTIC Document.
%Type = Inproceedings
\bibitem[{Hu et~al.(2009)Hu, Landa \& Shing}]{ukp_hu_landa_shing_survey}
\bibinfo{author}{Hu, T.~C.}, \bibinfo{author}{Landa, L.}, \&
  \bibinfo{author}{Shing, M.-T.} (\bibinfo{year}{2009}).
\newblock \bibinfo{title}{The unbounded knapsack problem}.
\newblock In \bibinfo{editor}{W.~Cook}, \bibinfo{editor}{L.~Lovász}, \&
  \bibinfo{editor}{J.~Vygen} (Eds.), {\it \bibinfo{booktitle}{Research Trends
  in Combinatorial Optimization}\/} (pp. \bibinfo{pages}{201 -- 217}).
\newblock \bibinfo{publisher}{Springer, Berlin, Heidelberg}.
\newblock \DOIprefix\doi{10.1007/978-3-540-76796-1_10}.
%Type = Misc
\bibitem[{IBM(2018)}]{cplex}
\bibinfo{author}{IBM} (\bibinfo{year}{2018}).
\newblock \bibinfo{title}{Cplex user’s manual}.
\newblock \URLprefix \url{https://www.ibm.com/analytics/cplex-optimizer}.
%Type = Book
\bibitem[{Kellerer et~al.(2004)Kellerer, Pferschy \& Pisinger}]{book_ukp_2004}
\bibinfo{author}{Kellerer, H.}, \bibinfo{author}{Pferschy, U.}, \&
  \bibinfo{author}{Pisinger, D.} (\bibinfo{year}{2004}).
\newblock {\it \bibinfo{title}{Knapsack problems}\/}.
\newblock \bibinfo{publisher}{Spinger-Verlag, Berlin}.
\newblock \DOIprefix\doi{10.1007/978-3-540-24777-7}.
%Type = Techreport
\bibitem[{Landa(2004)}]{landa_sage}
\bibinfo{author}{Landa, L.} (\bibinfo{year}{2004}).
\newblock {\it \bibinfo{title}{Sage Algorithms for Knapsack Problem}\/}.
\newblock \bibinfo{type}{Technical Report} \bibinfo{number}{CS2004-0794}
  University of California at San Diego \bibinfo{address}{9500 Gilman Drive, La
  Jolla, CA 92093-0021}.
\newblock \URLprefix
  \url{http://csetechrep.ucsd.edu/Dienst/UI/2.0/Describe/ncstrl.ucsd_cse/CS2004-0794}.
%Type = Article
\bibitem[{Martello \& Toth(1977)}]{mtu1}
\bibinfo{author}{Martello, S.}, \& \bibinfo{author}{Toth, P.}
  (\bibinfo{year}{1977}).
\newblock \bibinfo{title}{Branch-and-bound algorithms for the solution of the
  general unidimensional knapsack problem}.
\newblock {\it \bibinfo{journal}{Advances in Operations Research,
  North-Holland, Amsterdam}\/},  (pp. \bibinfo{pages}{295--301}).
%Type = Article
\bibitem[{Martello \& Toth(1990)}]{mtu2}
\bibinfo{author}{Martello, S.}, \& \bibinfo{author}{Toth, P.}
  (\bibinfo{year}{1990}).
\newblock \bibinfo{title}{An exact algorithm for large unbounded knapsack
  problems}.
\newblock {\it \bibinfo{journal}{Operations research letters}\/},  {\it
  \bibinfo{volume}{9}\/}, \bibinfo{pages}{15--20}.
  \DOIprefix\doi{10.1016/0167-6377(90)90035-4}.
%Type = Book
\bibitem[{Pisinger(1994)}]{pisinger1994dominance}
\bibinfo{author}{Pisinger, D.} (\bibinfo{year}{1994}).
\newblock {\it \bibinfo{title}{Dominance relations in unbounded knapsack
  problems. DIKU report 94/33}\/}.
\newblock \bibinfo{publisher}{DIKU}.
%Type = Inproceedings
\bibitem[{Poirriez \& Andonov(1998)}]{ukp_new_results}
\bibinfo{author}{Poirriez, V.}, \& \bibinfo{author}{Andonov, R.}
  (\bibinfo{year}{1998}).
\newblock \bibinfo{title}{Unbounded knapsack problem: new results}.
\newblock In \bibinfo{editor}{R.~Battiti}, \& \bibinfo{editor}{A.~A. Bertossi}
  (Eds.), {\it \bibinfo{booktitle}{Workshop Algorithms and Experiments,
  ALEX98}\/} (pp. \bibinfo{pages}{103--111}).
%Type = Article
\bibitem[{Poirriez et~al.(2009)Poirriez, Yanev \& Andonov}]{pya}
\bibinfo{author}{Poirriez, V.}, \bibinfo{author}{Yanev, N.}, \&
  \bibinfo{author}{Andonov, R.} (\bibinfo{year}{2009}).
\newblock \bibinfo{title}{A hybrid algorithm for the unbounded knapsack
  problem}.
\newblock {\it \bibinfo{journal}{Discrete Optimization}\/},  {\it
  \bibinfo{volume}{6}\/}, \bibinfo{pages}{110--124}.
  \DOIprefix\doi{10.1016/j.disopt.2008.09.004}.
%Type = Article
\bibitem[{Shapiro \& Wagner(1967)}]{turnpike}
\bibinfo{author}{Shapiro, J.~F.}, \& \bibinfo{author}{Wagner, H.~M.}
  (\bibinfo{year}{1967}).
\newblock \bibinfo{title}{A finite renewal algorithm for the knapsack and
  turnpike models}.
\newblock {\it \bibinfo{journal}{Operations Research}\/},  {\it
  \bibinfo{volume}{15}\/}, \bibinfo{pages}{319--341}.
  \DOIprefix\doi{10.1287/opre.15.2.319}.
%Type = Article
\bibitem[{Zhu \& Broughan(1997)}]{zhu_dominated}
\bibinfo{author}{Zhu, N.}, \& \bibinfo{author}{Broughan, K.}
  (\bibinfo{year}{1997}).
\newblock \bibinfo{title}{On dominated terms in the general knapsack problem}.
\newblock {\it \bibinfo{journal}{Operations Research Letters}\/},  {\it
  \bibinfo{volume}{21}\/}, \bibinfo{pages}{31--37}.
  \DOIprefix\doi{10.1016/S0167-6377(97)00018-7}.

\end{thebibliography}
\bibliographystyle{model5-names}
\biboptions{authoryear}

% Commented for paper generation, the appendix is a separate file.
\appendix

\section{Supplementary Materials -- Summary Tables}
\setcounter{figure}{0} % Make the first appendix table to be A.1

\begin{table}[H]
\label{tab:times_pya}
\def\arraystretch{1.1}
\setlength\tabcolsep{4px}

\begin{adjustbox}{max width=\textwidth, center}
\begin{tabular}{@{\extracolsep{4pt}}rrrrrrrrrrrr@{}}

\hline
\multicolumn{3}{l}{Instance desc.} & \multicolumn{3}{l}{R-TSO} & \multicolumn{3}{l}{R-GFDP} & \multicolumn{3}{l}{PYAsUKP}\\
\cline{1-3}\cline{4-6}\cline{7-9}\cline{10-12}

\multicolumn{3}{l}{400 inst. per line} & \multicolumn{9}{l}{Subset-sum. Random \emph{c} between \([5\times10^6; 10^7]\)}\\
\cline{1-3}\cline{4-12}

& \textbf{n} & \(w_{min}\)  & \textbf{avg} & \textbf{sd} & \textbf{max} & \textbf{avg} & \textbf{sd} & \textbf{max} & \textbf{avg} & \textbf{sd} & \textbf{max}\\
\cline{1-3}\cline{4-6}\cline{7-9}\cline{10-12}

\multicolumn{3}{c}{See caption.} & \textbf{0.05} & 0.12 & 0.75 & -- & -- & -- & 2.52 & 21.76 & 302.51\\ % & 0.05 & 0.12 & 0.74 & -- & -- & -- & 2.52 & 21.75 & 302.51 \\
\hline

\multicolumn{3}{l}{20 inst. per line} & \multicolumn{9}{l}{Strong correlation. Random \emph{c} between \([20\overline{n}; 100\overline{n}]\)}\\
\cline{1-3}\cline{4-12}
\textbf{\(\alpha\)} & \textbf{n} & \(w_{min}\) & \textbf{avg} & \textbf{sd} & \textbf{max} & \textbf{avg} & \textbf{sd} & \textbf{max} & \textbf{avg} & \textbf{sd} & \textbf{max}\\

\cline{1-3}\cline{4-6}\cline{7-9}\cline{10-12}
 5 & 5  & 10 & \textbf{0.04} & 0.00 & 0.04 & 0.04 & 0.00 & 0.05 & 1.57 & 1.78 & 3.62\\
   &    & 15 & \textbf{0.04} & 0.00 & 0.04 & 0.04 & 0.00 & 0.05 & 1.57 & 1.78 & 3.62\\%0.04 & 0.00 & 0.05 &  0.43 & 0.06 &  0.53 & 3.85 & 1.53 & 5.13\\
   &    & 50 & \textbf{0.04} & 0.00 & 0.04 & 0.04 & 0.00 & 0.05 & 1.57 & 1.78 & 3.62\\%0.13 & 0.00 & 0.16 &  1.01 & 0.52 &  1.70 & 12.12 & 8.17 & 28.84\\
 5 & 10 & 10 & \textbf{0.04} & 0.00 & 0.04 & 0.04 & 0.00 & 0.05 & 1.57 & 1.78 & 3.62\\%0.06 & 0.00 & 0.06 &  0.50 & 0.04 &  0.54 & 0.00 & 0.00 & 0.01\\
   &    & 50 & \textbf{0.04} & 0.00 & 0.04 & 0.04 & 0.00 & 0.05 & 1.57 & 1.78 & 3.62\\%0.29 & 0.00 & 0.30 &  5.93 & 0.82 &  6.79 & 22.43 & 17.85 & 45.19\\
   &    & 110& \textbf{0.04} & 0.00 & 0.04 & 0.04 & 0.00 & 0.05 & 1.57 & 1.78 & 3.62\\%0.66 & 0.00 & 0.66 & 16.05 & 3.36 & 19.68 & 76.53 & 62.54 & 175.61\\
-5 & 5  & 10 & \textbf{0.04} & 0.00 & 0.05 & 0.05 & 0.00 & 0.05 & 4.02 & 2.72 & 7.12\\%0.04 & 0.00 & 0.05 & 0.04 & 0.00 & 0.04 & 4.02 & 2.72 & 7.12\\
   &    & 15 & \textbf{0.04} & 0.00 & 0.05 & 0.05 & 0.00 & 0.05 & 4.02 & 2.72 & 7.12\\%0.05 & 0.00 & 0.05 & 0.05 & 0.00 & 0.05 & 6.76 & 4.22 & 12.24\\
   &    & 50 & \textbf{0.04} & 0.00 & 0.05 & 0.05 & 0.00 & 0.05 & 4.02 & 2.72 & 7.12\\%0.14 & 0.00 & 0.15 & 0.11 & 0.02 & 0.12 & 24.76 & 19.41 & 66.23\\
-5 & 10 & 10 & \textbf{0.12} & 0.00 & 0.12 & 0.15 & 0.01 & 0.16 & 6.74 & 6.28 & 15.38\\%0.10 & 0.00 & 0.10 & 0.11 & 0.01 & 0.13 & 6.74 & 6.28 & 15.38\\
   &    & 50 & \textbf{0.12} & 0.00 & 0.12 & 0.15 & 0.01 & 0.16 & 6.74 & 6.28 & 15.38\\%0.32 & 0.00 & 0.32 & 0.28 & 0.01 & 0.29 & 48.70 & 42.53 & 111.61\\
   &    & 110& \textbf{0.12} & 0.00 & 0.12 & 0.15 & 0.01 & 0.16 & 6.74 & 6.28 & 15.38\\%0.65 & 0.00 & 0.66 & 0.52 & 0.01 & 0.53 & 144.87 & 143.53 & 416.41\\
\hline

\multicolumn{3}{l}{200 inst. per line} & \multicolumn{9}{l}{Postponed periodicity. Random \emph{c} between \([w_{max}; 2\times10^6]\)}\\
\cline{1-3}\cline{4-12}
& \textbf{n} & \(w_{min}\) & \textbf{avg} & \textbf{sd} & \textbf{max} & \textbf{avg} & \textbf{sd} & \textbf{max} & \textbf{avg} & \textbf{sd} & \textbf{max}\\
\cline{1-3}\cline{4-6}\cline{7-9}\cline{10-12}
& 20 & 20 & \textbf{0.76} & 0.12 & 0.99 & 0.79 & 0.12 & 1.04 & 8.65 & 7.74 & 28.63\\%0.79 & 0.10 & 0.97 & 0.74 & 0.11 & 0.96 & 8.65 & 7.74 & 28.63\\
& 50 & 20 & \textbf{5.15} & 0.67 & 6.22 & 5.36 & 0.71 & 6.45 & 78.34 & 82.46 & 356.67\\%5.70 & 0.37 & 6.54 & 5.12 & 0.65 & 6.13 & 78.34 & 82.46 & 356.67\\
& 20 & 50 & \textbf{0.76} & 0.15 & 1.12 & 0.79 & 0.16 & 1.17 & 11.57 & 8.20 & 39.20\\%0.89 & 0.12 & 1.19 & 0.75 & 0.14 & 1.09 & 11.57 & 8.20 & 39.20\\
& 50 & 50 & \textbf{3.95} & 0.76 & 5.33 & 4.13 & 0.81 & 5.50 & 113.21 & 87.16 & 267.10\\%4.72 & 0.69 & 6.27 & 3.97 & 0.75 & 5.30 & 113.21 & 87.16 & 267.10\\
\hline

\multicolumn{3}{l}{500 inst. per line} & \multicolumn{9}{l}{No collective dominance. Random \emph{c} between \([w_{max}; 1000\overline{n}]\)}\\
\cline{1-3}\cline{4-12}
& \textbf{n} & \(w_{min}\) & \textbf{avg} & \textbf{sd} & \textbf{max} & \textbf{avg} & \textbf{sd} & \textbf{max} & \textbf{avg} & \textbf{sd} & \textbf{max}\\
\cline{1-3}\cline{4-6}\cline{7-9}\cline{10-12}
&  5 & n & 0.05 & 0.01 & 0.08 & \textbf{0.04} & 0.01 & 0.08 & 0.59 & 0.44 & 2.03\\% 0.07 & 0.03 & 0.14 & 0.04 & 0.01 & 0.07 & 0.59 & 0.44 & 2.03\\
& 10 & n & 0.41 & 0.15 & 0.85 & \textbf{0.34} & 0.10 & 0.63 & 2.34 & 1.86 & 8.44\\% 0.65 & 0.31 & 1.30 & 0.33 & 0.10 & 0.60 & 2.34 & 1.86 & 8.44\\
& 20 & n & 0.81 & 0.17 & 1.43 & \textbf{0.76} & 0.13 & 1.42 & 8.62 & 7.64 & 31.22\\% 1.04 & 0.32 & 1.91 & 0.72 & 0.12 & 1.31 & 8.62 & 7.64 & 31.22\\
& 50 & n & \textbf{3.71} & 0.26 & 4.93 & 3.72 & 0.21 & 4.85 & 73.49 & 72.26 & 279.01\\% 3.64 & 0.36 & 4.74 & 3.56 & 0.20 & 4.46 & 73.49 & 72.26 & 279.01\\
\hline

\multicolumn{3}{l}{\emph{qtd} inst. per line} & \multicolumn{9}{l}{SAW. Random \emph{c} between \([w_{max}; 10\overline{n}]\)}\\
\cline{1-3}\cline{4-12}
\textbf{qtd} & \textbf{n} & \(w_{min}\) & \textbf{avg} & \textbf{sd} & \textbf{max} & \textbf{avg} & \textbf{sd} & \textbf{max} & \textbf{avg} & \textbf{sd} & \textbf{max}\\
\cline{1-3}\cline{4-6}\cline{7-9}\cline{10-12}
~200 &  10 & 10 & \textbf{0.08} & 0.01 & 0.12 & 0.17 & 0.02 & 0.21 & 1.32 & 0.85 & 3.01 \\%0.08 & 0.00 & 0.09 & 0.14 & 0.02 & 0.21 & 1.32 & 0.85 & 3.01\\
~500 &  50 &  5 & \textbf{0.54} & 0.02 & 0.58 & 1.43 & 0.54 & 2.67 & 3.36 & 2.86 & 11.16  \\%0.50 & 0.01 & 0.53 & 2.09 & 1.00 & 3.75 & 3.36 & 2.86 & 11.16\\
~200 &  50 & 10 & \textbf{0.77} & 0.01 & 0.80 & 1.50 & 0.51 & 2.58 & 6.99 & 5.81 & 23.04 \\%0.72 & 0.01 & 0.74 & 2.15 & 0.85 & 3.65 & 6.99 & 5.81 & 23.04\\
~200 & 100 & 10 & \textbf{8.39} & 0.36 & 9.12 & 31.52 & 5.27 & 36.91 & 40.43 & 35.13 & 118.28\\%7.34 & 0.32 & 8.09 & 33.93 & 6.94 & 43.40 & 40.43 & 35.13 & 118.28\\
\hline

\end{tabular}
\end{adjustbox}
\caption{Time (in seconds) for the PYAsUKP 4540 Instances (see Section \ref{sec:pya_exp}). Columns \textbf{n} and \(w_{min}\) values must be multiplied by \(10^3\) to obtain their true value. Let \(T\) be the set of run times reported by the R-TSO, R-GFDP or EDUK2 for the instance dataset described by a row. The meaning of the columns \textbf{avg}, \textbf{sd} and \textbf{max}, is, respectively, the arithmetic mean of \(T\), the standard deviation of \(T\), the maximum value of \(T\). The notation \(x\overline{n}\) means \(x\) concatenated to the value of \(n\) (e.g. \(n = 5000\) then \(10\overline{n} = 105000\)). For the subset-sum instances, there are ten instances for each possible combination of: \(w_{min} \in \{10^3, 5\times10^3, 10^4, 5\times10^4, 10^5\}\); \(w_{max} \in \{5\times10^5, 10^6\}\) and \(n \in \{10^3, 2\times10^3, 5\times10^3, 10^4\}\), totaling 400 instances.}
\end{table}

\clearpage

\begin{table}[H]
\label{tab:times_mtu}
\vspace{1mm}
\def\arraystretch{1.1}
\setlength\tabcolsep{6px}

\begin{adjustbox}{max width=\textwidth, center}
\begin{tabular}{@{\extracolsep{4pt}}rrrrrrrrrrr@{}}

\hline
\multicolumn{3}{l}{Instance desc.} & \multicolumn{2}{c}{F77-MTU1} & \multicolumn{2}{c}{CPP-MTU1} & \multicolumn{2}{c}{F77-MTU2} & \multicolumn{2}{c}{CPP-MTU2}\\
\cline{1-3}\cline{4-5}\cline{6-7}\cline{8-9}\cline{10-11}

\multicolumn{3}{l}{40 inst. per line} & \multicolumn{8}{l}{Subset-sum. Random \emph{c} between \([5\times10^6; 10^7]\)}\\
\cline{1-3}\cline{4-11}

& \textbf{n} & \(w_{min}\) & \textbf{avg} & \textbf{fin}  & \textbf{avg} & \textbf{fin} & \textbf{avg} & \textbf{fin} & \textbf{avg} & \textbf{fin}\\
\cline{1-3}\cline{4-5}\cline{6-7}\cline{8-9}\cline{10-11}

\multicolumn{3}{c}{See caption.} & 0.01 & 40 & 0.04 & 40 & 154.93 & 8 & 0.04 & 40\\ %0.04 & 40 & 0.04 & 40 & 0.00 & 40 & 154.97 & 8\\
\hline

\multicolumn{3}{l}{2 inst. per line} & \multicolumn{8}{l}{Strong correlation. Random \emph{c} between \([20\overline{n}; 100\overline{n}]\)}\\
\cline{1-3}\cline{4-11}
\textbf{\(\alpha\)} & \textbf{n} & \(w_{min}\) & \textbf{avg} & \textbf{fin}  & \textbf{avg} & \textbf{fin} & \textbf{avg} & \textbf{fin} & \textbf{avg} & \textbf{fin}\\

\cline{1-3}\cline{4-5}\cline{6-7}\cline{8-9}\cline{10-11}
 5 & 5  & 10 & 0.00 & 1 & 0.00 & 1 & -- & 0 & -- & 0\\%0.00 & 1 &   -- & 0 & 0.00 & 1 &   -- & 0\\
   &    & 15 & -- & 0 & -- & 0 & -- & 0 & -- & 0\\%  -- & 0 &   -- & 0 &   -- & 0 &   -- & 0\\
   &    & 50 & -- & 0 & -- & 0 & -- & 0 & -- & 0\\%  -- & 0 &   -- & 0 &   -- & 0 &   -- & 0\\
 5 & 10 & 10 & 0.00 & 1 & 0.00 & 1 & -- & 0 & -- & 0\\%0.00 & 1 &   -- & 0 & 0.00 & 1 &   -- & 0\\
   &    & 50 & 0.03 & 1 & 0.05 & 1 & -- & 0 & -- & 0\\%0.04 & 1 &   -- & 0 & 0.03 & 1 &   -- & 0\\
   &    & 110& 0.01 & 1 & 0.01 & 1 & -- & 0 & -- & 0\\%0.01 & 1 &   -- & 0 & 0.00 & 1 &   -- & 0\\
-5 & 5  & 10 & -- & 0 & -- & 0 & -- & 0 & -- & 0\\%  -- & 0 &   -- & 0 &   -- & 0 &   -- & 0\\
   &    & 15 & -- & 0 & -- & 0 & -- & 0 & -- & 0\\%  -- & 0 &   -- & 0 &   -- & 0 &   -- & 0\\
   &    & 50 & -- & 0 & -- & 0 & -- & 0 & -- & 0\\%  -- & 0 &   -- & 0 &   -- & 0 &   -- & 0\\
-5 & 10 & 10 & 0.00 & 1 & 0.00 & 1 & -- & 0 & -- & 0\\%0.00 & 1 &   -- & 0 & 0.00 & 1 &   -- & 0\\
   &    & 50 & 0.00 & 1 & 0.00 & 1 & 0.83 & 1 & 0.79 & 1\\%0.00 & 1 & 0.79 & 1 & 0.00 & 1 & 0.83 & 1\\
   &    & 110& 0.00 & 1 & 0.00 & 1 & -- & 0 & -- & 0\\%0.00 & 1 &   -- & 0 & 0.00 & 1 &   -- & 0\\
\hline

\multicolumn{3}{l}{20 inst. per line} & \multicolumn{8}{l}{Postponed periodicity. Random \emph{c} between \([w_{max}; 2\times10^6]\)}\\
\cline{1-3}\cline{4-11}
& \textbf{n} & \(w_{min}\) & \textbf{avg} & \textbf{fin}  & \textbf{avg} & \textbf{fin} & \textbf{avg} & \textbf{fin} & \textbf{avg} & \textbf{fin}\\
\cline{1-3}\cline{4-5}\cline{6-7}\cline{8-9}\cline{10-11}
& 20 & 20 & 134.10 & 19 & 67.16 & 19 & 74.44 & 18 & 33.26 & 18\\%67.17 & 19 & 67.17 & 19 &  18.05 & 17 &  15.12 & 17\\
& 50 & 20 & 204.95 & 18 & 257.23 & 20 & 214.30 & 18 & 2.15 & 20\\% 1.93 & 18 &  2.15 & 20 & 134.09 & 17 & 143.47 & 17\\
& 20 & 50 & 6.33 & 20 & 3.15 & 20 & 7.83 & 20 & 1.74 & 20\\% 3.15 & 20 &  1.74 & 20 &   6.33 & 20 &   7.83 & 20\\
& 50 & 50 & 4.45 & 20 & 2.22 & 20 & 13.81 & 20 & 21.13 & 20\\% 2.22 & 20 & 21.13 & 20 &   4.45 & 20 &  13.81 & 20\\
\hline

\multicolumn{3}{l}{50 inst. per line} & \multicolumn{8}{l}{No collective dominance. Random \emph{c} between \([w_{max}; 1000\overline{n}]\)}\\
\cline{1-3}\cline{4-11}
& \textbf{n} & \(w_{min}\) & \textbf{avg} & \textbf{fin} & \textbf{avg} & \textbf{fin} & \textbf{avg} & \textbf{fin} & \textbf{avg} & \textbf{fin}\\
\cline{1-3}\cline{4-5}\cline{6-7}\cline{8-9}\cline{10-11}
&  5 & n & 19.68 & 9 & 16.54 & 9 & 37.38 & 9 & 36.93 & 9  \\% 16.54 & 9 & 37.01 & 9 & 19.85 & 9 & 37.29 & 9\\
& 10 & n & 276.36 & 6 & 147.01 & 6 & 308.40 & 6 & 301.69 & 6\\%147.08 & 6 &  5.84 & 5 & 34.41 & 5 & 10.09 & 5\\
& 20 & n & 27.42 & 3 & 17.98 & 3 & 27.78 & 3 & 19.22 & 3\\% 17.95 & 3 & 19.23 & 3 & 27.45 & 3 & 27.78 & 3\\
& 50 & n & 26.73 & 2 & 513.12 & 3 & 2.64 & 2 & 1.40 & 2\\% 13.36 & 2 &  1.40 & 2 & 26.73 & 2 &  2.64 & 2\\
\hline

\multicolumn{3}{l}{\emph{qtd} inst. per line} & \multicolumn{8}{l}{SAW. Random \emph{c} between \([w_{max}; 10\overline{n}]\)}\\
\cline{1-3}\cline{4-11}
\textbf{qtd} & \textbf{n} & \(w_{min}\) & \textbf{avg} & \textbf{fin}  & \textbf{avg} & \textbf{fin} & \textbf{avg} & \textbf{fin} & \textbf{avg} & \textbf{fin}\\
\cline{1-3}\cline{4-5}\cline{6-7}\cline{8-9}\cline{10-11}
~20 &  10 & 10 & 2.54 & 20 & 1.33 & 20 & 2.87 & 20 & 12.92 & 20\\% 1.33 & 20 & 12.92 & 20 &  2.54 & 20 &  2.87 & 20\\
~50 &  50 &  5 & 86.05 & 46 & 70.10 & 47 & 85.76 & 42 & 101.79 & 39 \\%43.08 & 46 & 43.97 & 19 & 59.14 & 38 & 38.63 & 16\\
~20 &  50 & 10 & 87.67 & 19 & 43.97 & 19 & 85.97 & 19 & 38.63 & 16\\%55.14 & 45 & 87.68 & 19 & 47.05 & 41 & 85.97 & 19\\
~20 & 100 & 10 & 20.22 & 16 & 10.10 & 16 & 44.40 & 16 & 38.42 & 17\\%10.10 & 16 & 38.41 & 17 & 20.23 & 16 & 44.41 & 16\\
\hline
\end{tabular}
\end{adjustbox}
\caption{Time in seconds for the MTU implementations over the reduced PYAsUKP dataset (see Section \ref{sec:mtu_exp}). Columns \textbf{n} and \(w_{min}\) values must be multiplied by \(10^3\) to obtain their true value. Let \(T\) be the set of run times reported by CPP-MTU1, CPP-MTU2, F77-MTU1 and F77-MTU2, for the instance dataset described by a row (in this case, we do not count runs that ended in timeout). The meaning of the columns \textbf{avg} and \textbf{fin}ished, is, respectively, the arithmetic mean of \(T\) and the cardinality of \(T\). The time limit was set to 30 minutes. The notation \(x\overline{n}\) means \(x\) concatenated to the value of \(n\) (e.g. \(n = 5000\) then \(10\overline{n} = 105000\)). For the subset-sum instances, there is one instance for each possible combination of: \(w_{min} \in \{10^3, 5\times10^3, 10^4, 5\times10^4, 10^5\}\); \(w_{max} \in \{5\times10^5, 10^6\}\) and \(n \in \{10^3, 2\times10^3, 5\times10^3, 10^4\}\), totaling 40 instances.}
\end{table}

\begin{table}[H]
\label{tab:times_solvers}
\vspace{1mm}
\def\arraystretch{1.09}
\setlength\tabcolsep{6px}

%\begin{adjustbox}{max width=\textwidth, center}
\begin{tabular}{@{\extracolsep{4pt}}rrrrrrrrr@{}}

\hline
\multicolumn{3}{l}{Instance desc.} & \multicolumn{3}{c}{CPLEX} & \multicolumn{3}{c}{Gurobi}\\
\cline{1-3}\cline{4-6}\cline{7-9}

\multicolumn{3}{l}{40 inst. per line} & \multicolumn{6}{l}{Subset-sum. Random \(c \in [5\times10^6; 10^7]\)}\\
\cline{1-3}\cline{4-9}

& \textbf{n} & \(w_{min}\) & \textbf{avg} & \textbf{sd}  & \textbf{fin} & \textbf{avg} & \textbf{sd} & \textbf{fin}\\
\cline{1-3}\cline{4-6}\cline{7-9}

\multicolumn{3}{c}{See caption.} & 8.61 & 23.14 & 40 & 17.50 & 40.28 & 39\\
\hline

\multicolumn{3}{l}{2 inst. per line} & \multicolumn{6}{l}{Strong correlation. Random \(c \in [20\overline{n}; 100\overline{n}]\)}\\
\cline{1-3}\cline{4-9}
\textbf{\(\alpha\)} & \textbf{n} & \(w_{min}\) & \textbf{avg} & \textbf{sd}  & \textbf{fin} & \textbf{avg} & \textbf{sd} & \textbf{fin}\\

\cline{1-3}\cline{4-6}\cline{7-9}
 5 & 5  & 10 & 0.14 & 0.14 & 2 & 0.04 & 0.00 & 2\\
   &    & 15 & 0.04 & 0.00 & 2 & 0.04 & 0.00 & 2\\
   &    & 50 & 0.04 & 0.00 & 2 & 0.17 & 0.14 & 2\\
 5 & 10 & 10 & 0.44 & 0.53 & 2 & 0.09 & 0.05 & 2\\
   &    & 50 & 0.08 & 0.00 & 2 & 2.91 & 3.53 & 2\\
   &    & 110& 0.17 & 0.13 & 2 & 2.90 & 3.17 & 2\\
-5 & 5  & 10 & 0.04 & 0.01 & 2 & 0.04 & 0.00 & 2\\
   &    & 15 & 0.04 & 0.00 & 2 & 0.04 & 0.00 & 2\\
   &    & 50 & 0.05 & 0.02 & 2 & 0.27 & 0.30 & 2\\
-5 & 10 & 10 & 0.08 & 0.03 & 2 & 0.06 & 0.02 & 2\\
   &    & 50 & 0.11 & 0.09 & 2 & 1.82 & 2.53 & 2\\
   &    & 110& 0.08 & 0.00 & 2 & -- & -- & 0\\
\hline

\multicolumn{3}{l}{20 inst. per line} & \multicolumn{6}{l}{Postponed periodicity. Random \(c \in [w_{max}; 2\times10^6]\)}\\
\cline{1-3}\cline{4-9}
& \textbf{n} & \(w_{min}\) & \textbf{avg} & \textbf{sd} & \textbf{fin} & \textbf{avg} & \textbf{sd} & \textbf{fin}\\
\cline{1-3}\cline{4-6}\cline{7-9}
& 20 & 20 & 173.52 & 502.95 & 20 & 11.56 & 17.01 & 19\\
& 50 & 20 & 208.02 & 562.58 & 19 & 302.38 & 530.97 & 19\\
& 20 & 50 & 111.42 & 400.92 & 20 & 24.09 & 47.27 & 20\\
& 50 & 50 & 466.38 & 636.74 & 16 & 424.27 & 399.85 & 20\\
\hline

\multicolumn{3}{l}{50 inst. per line} & \multicolumn{6}{l}{No collective dominance. Random \(c \in [w_{max}; 1000\overline{n}]\)}\\
\cline{1-3}\cline{4-9}
& \textbf{n} & \(w_{min}\) & \textbf{avg} & \textbf{sd} & \textbf{fin} & \textbf{avg} & \textbf{sd} & \textbf{fin}\\
\cline{1-3}\cline{4-6}\cline{7-9}
&  5 & n & 36.29 & 255.57 & 50 & 0.29 & 0.38 & 49\\
& 10 & n & 0.13 & 0.08 & 50 & 3.71 & 23.30 & 47\\
& 20 & n & 0.26 & 0.29 & 50 & 172.77 & 405.94 & 47\\
& 50 & n & 181.13 & 546.97 & 50 & 284.14 & 643.39 & 23\\
\hline

\multicolumn{3}{l}{\emph{qtd} inst. per line} & \multicolumn{6}{l}{SAW. Random \(c \in [w_{max}; 10\overline{n}]\)}\\
\cline{1-3}\cline{4-9}
\textbf{qtd} & \textbf{n} & \(w_{min}\) & \textbf{avg} & \textbf{sd}  & \textbf{fin} & \textbf{avg} & \textbf{sd} & \textbf{fin}\\
\cline{1-3}\cline{4-6}\cline{7-9}
~20 &  10 & 10 & 9.56 & 14.69 & 20 & 13.72 & 18.05 & 20\\
~50 &  50 &  5 & 694.39 & 802.98 & 50 & 544.64 & 643.32 & 19\\
~20 &  50 & 10 & 675.54 & 812.98 & 19 & 677.55 & 689.23 & 10\\
~20 & 100 & 10 & 692.25 & 777.33 & 20 & 1061.34 & 815.67 & 8\\
\hline
\end{tabular}
%\end{adjustbox}
\caption{Time in seconds for the CPLEX and Gurobi over the reduced PYAsUKP dataset (see Section \ref{sec:cplex_gurobi_exp}). Columns \textbf{n} and \(w_{min}\) values must be multiplied by \(10^3\) to obtain their true value. Let \(T\) be the set of run times (not considering timeouts, memory exhaustion or wrong answer) for the instance dataset described by a row. The meaning of the columns \textbf{avg}, \textbf{sd} and \textbf{fin}ished, is, respectively, the arithmetic mean of \(T\), the standard deviation of \(T\), and the cardinality of \(T\). The time limit was set to 30 minutes. The notation \(x\overline{n}\) means \(x\) concatenated to the value of \(n\) (e.g. \(n = 5000\) then \(10\overline{n} = 105000\)). For the subset-sum instances, there is one instance for each possible combination of: \(w_{min} \in \{10^3, 5\times10^3, 10^4, 5\times10^4, 10^5\}\); \(w_{max} \in \{5\times10^5, 10^6\}\) and \(n \in \{10^3, 2\times10^3, 5\times10^3, 10^4\}\), totaling 40 instances.}
\end{table}

\clearpage

% latex table generated in R 3.5.3 by xtable 1.8-3 package
% Wed Mar 20 19:27:41 2019
\begin{longtable}{crrrrrc}
  \hline
algorithm & n & avg & sd & min & max & fin \\ 
  \hline \endhead  \hline
CPLEX & 1024 & 0.02 & 0.01 & 0.01 & 0.06 &  10 \\ 
  CPLEX & 2048 & 0.07 & 0.12 & 0.02 & 0.41 &  10 \\ 
  CPLEX & 4096 & 0.32 & 0.58 & 0.04 & 1.93 &  10 \\ 
  CPLEX & 8192 & 2.52 & 7.48 & 0.06 & 23.79 &  10 \\ 
  CPLEX & 16384 & 102.29 & 317.61 & 0.14 & 1006.17 &  10 \\ 
  CPLEX & 32768 & 1.31 & 1.35 & 0.30 & 4.22 &  10 \\ 
  CPLEX & 65536 & 20.72 & 36.62 & 1.24 & 114.24 &   9 \\ 
  CPLEX & 131072 & 216.80 & 513.31 & 1.35 & 1264.53 &   6 \\ 
  MTU1 (C++) & 1024 & 0.00 & 0.01 & 0.00 & 0.02 &  10 \\ 
  MTU1 (C++) & 2048 & 0.01 & 0.03 & 0.00 & 0.09 &  10 \\ 
  MTU1 (C++) & 4096 & 0.07 & 0.15 & 0.00 & 0.49 &  10 \\ 
  MTU1 (C++) & 8192 & 42.39 & 130.63 & 0.00 & 414.11 &  10 \\ 
  MTU1 (C++) & 16384 & 3.20 & 6.60 & 0.00 & 17.74 &  10 \\ 
  MTU1 (C++) & 32768 & 146.97 & 314.26 & 0.00 & 709.01 &   5 \\ 
  MTU1 (C++) & 65536 & 6.63 & 8.50 & 0.62 & 12.64 &   2 \\ 
  MTU1 (C++) & 131072 & 2.58 & -- & 2.58 & 2.58 &   1 \\ 
  MTU2 (C++) & 1024 & 0.00 & 0.00 & 0.00 & 0.00 &  10 \\ 
  MTU2 (C++) & 2048 & 0.68 & 1.96 & 0.00 & 6.24 &  10 \\ 
  MTU2 (C++) & 4096 & 1.44 & 3.05 & 0.00 & 9.42 &  10 \\ 
  MTU2 (C++) & 8192 & 53.35 & 104.74 & 0.00 & 295.46 &   9 \\ 
  MTU2 (C++) & 16384 & 4.58 & 13.61 & 0.00 & 40.89 &   9 \\ 
  MTU2 (C++) & 32768 & 22.97 & 33.53 & 0.00 & 71.32 &   4 \\ 
  MTU2 (C++) & 65536 & 0.85 & 0.71 & 0.35 & 1.35 &   2 \\ 
  MTU2 (C++) & 131072 & 2.27 & -- & 2.27 & 2.27 &   1 \\ 
  EDUK & 1024 & 0.01 & 0.00 & 0.00 & 0.01 &  10 \\ 
  EDUK & 2048 & 0.02 & 0.01 & 0.01 & 0.04 &  10 \\ 
  EDUK & 4096 & 0.07 & 0.08 & 0.02 & 0.29 &  10 \\ 
  EDUK & 8192 & 0.19 & 0.29 & 0.05 & 0.99 &  10 \\ 
  EDUK & 16384 & 0.79 & 0.82 & 0.21 & 2.52 &  10 \\ 
  EDUK & 32768 & 3.50 & 4.56 & 0.83 & 15.63 &  10 \\ 
  EDUK & 65536 & 26.56 & 31.51 & 3.68 & 111.18 &  10 \\ 
  EDUK & 131072 & 120.63 & 132.12 & 13.92 & 397.16 &  10 \\ 
  EDUK2 & 1024 & 0.00 & 0.00 & 0.00 & 0.00 &  10 \\ 
  EDUK2 & 2048 & 0.02 & 0.03 & 0.00 & 0.09 &  10 \\ 
  EDUK2 & 4096 & 0.05 & 0.07 & 0.01 & 0.24 &  10 \\ 
  EDUK2 & 8192 & 0.08 & 0.16 & 0.00 & 0.53 &  10 \\ 
  EDUK2 & 16384 & 0.29 & 0.55 & 0.02 & 1.75 &  10 \\ 
  EDUK2 & 32768 & 1.13 & 1.35 & 0.02 & 4.35 &  10 \\ 
  EDUK2 & 65536 & 15.84 & 28.41 & 0.07 & 94.79 &  10 \\ 
  EDUK2 & 131072 & 50.69 & 75.98 & 0.19 & 238.33 &  10 \\ 
  R-GFDP & 1024 & 0.00 & 0.00 & 0.00 & 0.01 &  10 \\ 
  R-GFDP & 2048 & 0.02 & 0.01 & 0.01 & 0.02 &  10 \\ 
  R-GFDP & 4096 & 0.04 & 0.02 & 0.02 & 0.10 &  10 \\ 
  R-GFDP & 8192 & 0.13 & 0.07 & 0.04 & 0.31 &  10 \\ 
  R-GFDP & 16384 & 0.41 & 0.24 & 0.21 & 0.90 &  10 \\ 
  R-GFDP & 32768 & 1.31 & 1.39 & 0.25 & 4.90 &  10 \\ 
  R-GFDP & 65536 & 7.07 & 5.37 & 1.14 & 20.23 &  10 \\ 
  R-GFDP & 131072 & 33.59 & 30.89 & 5.99 & 100.92 &  10 \\ 
  R-OSO & 1024 & 0.01 & 0.00 & 0.01 & 0.02 &  10 \\ 
  R-OSO & 2048 & 0.02 & 0.00 & 0.02 & 0.03 &  10 \\ 
  R-OSO & 4096 & 0.06 & 0.02 & 0.05 & 0.12 &  10 \\ 
  R-OSO & 8192 & 0.17 & 0.07 & 0.09 & 0.34 &  10 \\ 
  R-OSO & 16384 & 0.49 & 0.24 & 0.28 & 0.97 &  10 \\ 
  R-OSO & 32768 & 1.44 & 1.38 & 0.45 & 5.02 &  10 \\ 
  R-OSO & 65536 & 7.30 & 5.34 & 1.41 & 20.38 &  10 \\ 
  R-OSO & 131072 & 33.89 & 30.77 & 6.43 & 100.86 &  10 \\ 
  R-TSO & 1024 & 0.01 & 0.00 & 0.00 & 0.02 &  10 \\ 
  R-TSO & 2048 & 0.02 & 0.01 & 0.01 & 0.03 &  10 \\ 
  R-TSO & 4096 & 0.06 & 0.02 & 0.04 & 0.12 &  10 \\ 
  R-TSO & 8192 & 0.16 & 0.07 & 0.06 & 0.34 &  10 \\ 
  R-TSO & 16384 & 0.47 & 0.24 & 0.28 & 0.97 &  10 \\ 
  R-TSO & 32768 & 1.41 & 1.38 & 0.31 & 4.97 &  10 \\ 
  R-TSO & 65536 & 7.26 & 5.29 & 1.41 & 20.19 &  10 \\ 
  R-TSO & 131072 & 33.52 & 30.47 & 6.43 & 99.91 &  10 \\ 
  \hline
\caption{Time (in seconds) for the experiment with realistic random dataset (see \autoref{sec:rr_exp}). For each row, there is a set \(T\) comprised by the run times that \textbf{algorithm} spent solving instances of size \textbf{n}. We do not count the run time of runs that ended in timeout. The meaning of the columns \textbf{avg}, \textbf{sd}, \textbf{min}, \textbf{max} and \textbf{fin}ished are, respectively, the arithmetic mean of \(T\), the standard deviation of \(T\), the minimal value in \(T\), the maximal value in \(T\), and the cardinality of \(T\). The time limit was set to 30 minutes.} 
\label{tab:rr}
\end{longtable}

\clearpage

% latex table generated in R 3.5.3 by xtable 1.8-3 package
% Wed Mar 20 19:27:42 2019
\begin{longtable}{crrrrrc}
  \hline
algorithm & n & avg & sd & min & max & fin \\ 
  \hline \endhead  \hline
CPLEX & 2048 & 0.02 & 0.00 & 0.02 & 0.03 &  10 \\ 
  CPLEX & 4096 & 0.04 & 0.00 & 0.04 & 0.05 &  10 \\ 
  CPLEX & 8192 & 0.07 & 0.00 & 0.07 & 0.08 &  10 \\ 
  CPLEX & 16384 & 0.16 & 0.01 & 0.15 & 0.19 &  10 \\ 
  CPLEX & 32768 & 0.35 & 0.02 & 0.33 & 0.38 &  10 \\ 
  CPLEX & 65536 & 0.79 & 0.12 & 0.72 & 1.11 &  10 \\ 
  CPLEX & 131072 & 1.86 & 0.27 & 1.73 & 2.61 &  10 \\ 
  CPLEX & 262144 & 5.81 & 3.49 & 4.31 & 15.55 &  10 \\ 
  CPLEX & 524288 & 101.86 & 69.56 & 10.49 & 249.18 &  10 \\ 
  CPLEX & 1048576 & 405.39 & 349.01 & 22.90 & 1115.63 &  10 \\ 
  MTU1 (C++) & 2048 & 0.00 & 0.00 & 0.00 & 0.00 &  10 \\ 
  MTU1 (C++) & 4096 & 0.00 & 0.00 & 0.00 & 0.00 &  10 \\ 
  MTU1 (C++) & 8192 & 0.00 & 0.00 & 0.00 & 0.00 &  10 \\ 
  MTU1 (C++) & 16384 & 0.00 & 0.00 & 0.00 & 0.00 &  10 \\ 
  MTU1 (C++) & 32768 & 0.00 & 0.00 & 0.00 & 0.00 &  10 \\ 
  MTU1 (C++) & 65536 & 0.00 & 0.00 & 0.00 & 0.01 &  10 \\ 
  MTU1 (C++) & 131072 & 0.01 & 0.00 & 0.01 & 0.01 &  10 \\ 
  MTU1 (C++) & 262144 & 0.02 & 0.00 & 0.02 & 0.02 &  10 \\ 
  MTU1 (C++) & 524288 & 0.05 & 0.00 & 0.04 & 0.05 &  10 \\ 
  MTU1 (C++) & 1048576 & 0.09 & 0.00 & 0.09 & 0.10 &  10 \\ 
  MTU2 (C++) & 2048 & 0.00 & 0.00 & 0.00 & 0.00 &  10 \\ 
  MTU2 (C++) & 4096 & 0.00 & 0.00 & 0.00 & 0.00 &  10 \\ 
  MTU2 (C++) & 8192 & 0.00 & 0.00 & 0.00 & 0.00 &  10 \\ 
  MTU2 (C++) & 16384 & 0.00 & 0.00 & 0.00 & 0.00 &  10 \\ 
  MTU2 (C++) & 32768 & 0.00 & 0.00 & 0.00 & 0.00 &  10 \\ 
  MTU2 (C++) & 65536 & 0.00 & 0.00 & 0.00 & 0.00 &  10 \\ 
  MTU2 (C++) & 131072 & 0.00 & 0.00 & 0.00 & 0.00 &  10 \\ 
  MTU2 (C++) & 262144 & 0.01 & 0.00 & 0.01 & 0.01 &  10 \\ 
  MTU2 (C++) & 524288 & 0.01 & 0.00 & 0.01 & 0.01 &  10 \\ 
  MTU2 (C++) & 1048576 & 0.03 & 0.00 & 0.02 & 0.03 &  10 \\ 
  EDUK & 2048 & 0.42 & 0.36 & 0.13 & 1.02 &  10 \\ 
  EDUK & 4096 & 2.69 & 3.29 & 0.69 & 11.34 &  10 \\ 
  EDUK & 8192 & 4.33 & 1.48 & 2.24 & 6.34 &  10 \\ 
  EDUK & 16384 & 53.39 & 81.66 & 13.40 & 271.78 &  10 \\ 
  EDUK & 32768 & 192.91 & 144.41 & 59.95 & 459.60 &  10 \\ 
  EDUK & 65536 & 798.05 & 342.29 & 463.72 & 1619.03 &   9 \\ 
  EDUK & 131072 & -- & -- & -- & -- &   0 \\ 
  EDUK & 262144 & -- & -- & -- & -- &   0 \\ 
  EDUK & 524288 & -- & -- & -- & -- &   0 \\ 
  EDUK & 1048576 & -- & -- & -- & -- &   0 \\ 
  EDUK2 & 2048 & 0.00 & 0.00 & 0.00 & 0.00 &  10 \\ 
  EDUK2 & 4096 & 0.00 & 0.00 & 0.00 & 0.00 &  10 \\ 
  EDUK2 & 8192 & 0.00 & 0.00 & 0.00 & 0.01 &  10 \\ 
  EDUK2 & 16384 & 0.01 & 0.00 & 0.01 & 0.01 &  10 \\ 
  EDUK2 & 32768 & 0.01 & 0.00 & 0.01 & 0.01 &  10 \\ 
  EDUK2 & 65536 & 0.03 & 0.00 & 0.02 & 0.03 &  10 \\ 
  EDUK2 & 131072 & 0.05 & 0.00 & 0.05 & 0.06 &  10 \\ 
  EDUK2 & 262144 & 0.11 & 0.00 & 0.10 & 0.12 &  10 \\ 
  EDUK2 & 524288 & 0.22 & 0.00 & 0.21 & 0.23 &  10 \\ 
  EDUK2 & 1048576 & 0.45 & 0.04 & 0.33 & 0.47 &  10 \\ 
  R-GFDP & 2048 & 0.01 & 0.01 & 0.00 & 0.03 &  10 \\ 
  R-GFDP & 4096 & 0.05 & 0.12 & 0.00 & 0.36 &  10 \\ 
  R-GFDP & 8192 & 0.02 & 0.06 & 0.00 & 0.19 &  10 \\ 
  R-GFDP & 16384 & 0.00 & 0.00 & 0.00 & 0.01 &  10 \\ 
  R-GFDP & 32768 & 2.68 & 8.45 & 0.01 & 26.72 &  10 \\ 
  R-GFDP & 65536 & 0.02 & 0.00 & 0.01 & 0.02 &  10 \\ 
  R-GFDP & 131072 & 0.03 & 0.01 & 0.03 & 0.04 &  10 \\ 
  R-GFDP & 262144 & 0.07 & 0.01 & 0.05 & 0.08 &  10 \\ 
  R-GFDP & 524288 & 0.14 & 0.02 & 0.10 & 0.16 &   9 \\ 
  R-GFDP & 1048576 & 0.27 & 0.05 & 0.21 & 0.32 &   9 \\ 
  R-OSO & 2048 & 0.01 & 0.01 & 0.01 & 0.03 &  10 \\ 
  R-OSO & 4096 & 0.08 & 0.10 & 0.02 & 0.35 &  10 \\ 
  R-OSO & 8192 & 0.13 & 0.04 & 0.08 & 0.19 &  10 \\ 
  R-OSO & 16384 & 2.95 & 5.39 & 0.47 & 17.40 &  10 \\ 
  R-OSO & 32768 & 8.90 & 10.22 & 1.78 & 28.67 &  10 \\ 
  R-OSO & 65536 & 30.59 & 39.70 & 8.41 & 127.06 &  10 \\ 
  R-OSO & 131072 & 90.47 & 83.21 & 32.84 & 281.02 &  10 \\ 
  R-OSO & 262144 & 442.53 & 316.66 & 154.98 & 1173.33 &  10 \\ 
  R-OSO & 524288 & 1167.81 & 523.10 & 588.01 & 1604.37 &   3 \\ 
  R-OSO & 1048576 & -- & -- & -- & -- &   0 \\ 
  R-TSO & 2048 & 0.02 & 0.02 & 0.01 & 0.07 &  10 \\ 
  R-TSO & 4096 & 0.08 & 0.10 & 0.02 & 0.35 &  10 \\ 
  R-TSO & 8192 & 0.13 & 0.04 & 0.08 & 0.20 &  10 \\ 
  R-TSO & 16384 & 2.95 & 5.41 & 0.48 & 17.50 &  10 \\ 
  R-TSO & 32768 & 8.91 & 10.28 & 1.77 & 28.74 &  10 \\ 
  R-TSO & 65536 & 30.49 & 39.47 & 8.27 & 126.18 &  10 \\ 
  R-TSO & 131072 & 90.29 & 83.18 & 32.78 & 280.90 &  10 \\ 
  R-TSO & 262144 & 442.31 & 316.89 & 154.69 & 1173.17 &  10 \\ 
  R-TSO & 524288 & 1171.31 & 524.86 & 589.44 & 1609.04 &   3 \\ 
  R-TSO & 1048576 & -- & -- & -- & -- &   0 \\ 
  \hline
\caption{Time (in seconds) for the BREQ 128-16 Standard Benchmark (see Section \ref{sec:breq_exp}). For each row, there is a set \(T\) comprised by the run times that \textbf{algorithm} spent solving instances of size \textbf{n}. We do not count the run time of runs that ended in timeout. The meaning of the columns \textbf{avg}, \textbf{sd}, \textbf{min}, \textbf{max} and \textbf{fin}ished are, respectively, the arithmetic mean of \(T\), the standard deviation of \(T\), the minimal value in \(T\), the maximal value in \(T\), and the cardinality of \(T\). The time limit was set to 30 minutes.} 
\label{tab:breq}
\end{longtable}

\begin{table}
\begin{adjustbox}{max width=\textwidth, center}
\begin{tabu} to \linewidth { lrrrrrr }
Dataset  & N & N' & \makecell{Master\\time} & \makecell{Pricing\\time} & \makecell{\# of sub-\\problems} & \makecell{Time per\\subproblem}\\\\\multicolumn{7}{c}{MTU1}\\
ANI\&AI  &  500 & 451 & 0.47 & 209.31 &  714.45 & 0.25912 \\
Falkenauer  &  160 &   0 & 0.08 & 0.04 &  464.81 & 0.00005 \\
GI  &  240 &   0 & 2.60 & 4.47 & 1755.90 & 0.00178 \\
Hard28  &   28 &   0 & 0.19 & 0.05 &  565.46 & 0.00009 \\
Random  & 3840 &   3 & 0.19 & 0.06 &  627.54 & 0.00003 \\
Scholl  & 1210 &   0 & 0.03 & 0.10 &  183.15 & 0.00014 \\
Schwerin  &  200 &   0 & 0.00 & 0.25 &   74.18 & 0.00330 \\
Waescher  &   17 &   0 & 0.01 & 0.01 &  135.82 & 0.00006 \\
\multicolumn{7}{c}{CPLEX}\\
ANI\&AI  &  500 & 491 & 0.33 & 1091.07 & 629.44 & 1.72159 \\
Falkenauer  &  160 &   0 & 0.07 & 4.90 & 358.52 & 0.01044 \\
GI  &  240 &  85 & 0.42 & 367.39 & 748.26 & 0.36982 \\
Hard28  &   28 &   0 & 0.19 & 30.68 & 538.71 & 0.05526 \\
Random  & 3840 &   3 & 0.13 & 18.33 & 366.79 & 0.01817 \\
Scholl  & 1210 &  10 & 0.03 & 7.30 & 148.80 & 0.03775 \\
Schwerin  &  200 &   0 & 0.01 & 197.23 &  70.95 & 2.77470 \\
Waescher  &   17 &   0 & 0.02 & 9.72 & 131.82 & 0.05485 \\
\multicolumn{7}{c}{Ordered Step-off (integer, no sort)}\\
ANI\&AI  &  500 & 0 & 26.01 & 8.23 & 2896.43 & 0.00183 \\
Falkenauer  &  160 & 0 & 0.08 & 0.01 &  467.11 & 0.00002 \\
GI  &  240 & 0 & 2.80 & 10.10 & 1743.95 & 0.00539 \\
Hard28  &   28 & 0 & 0.19 & 0.02 &  561.93 & 0.00004 \\
Random  & 3840 & 3 & 0.19 & 0.02 &  630.07 & 0.00001 \\
Scholl  & 1210 & 0 & 0.04 & 0.01 &  183.43 & 0.00002 \\
Schwerin  &  200 & 0 & 0.01 & 0.00 &   74.33 & 0.00002 \\
Waescher  &   17 & 0 & 0.01 & 0.02 &  135.76 & 0.00012 \\
\end{tabu}

\end{adjustbox}
\caption{Results of the experiment focused in solving pricing subproblems of the BPP/CSP (see \autoref{sec:csp_experiments}). The results of solving the pricing problems with CPLEX are included here for completeness. The time limit was 30 minutes for the total run time (which includes reading input, the master time and the pricing time). The unit of all time columns is seconds. The meaning of each column is: N -- amount of instances in the respective dataset; N' -- amount of instances in which the solver coupled with the respective algorithm exceeded the time limit; Master time -- amount of time taken by CPLEX to solve the linear programming relaxation (disregarding the solving of pricing problems); Pricing time -- mean time spent by the respective algorithm solving all pricing problems of a single instance of the respective dataset; \# of subproblems -- the mean amount of generated pricing problems in a single instance of the respective dataset (this amount is affected by the exact optimal solutions which were returned by the algorithm used to solve the pricing problems); Time per subproblem -- the result of the division of the two last columns. } 
\label{tab:csp}
\end{table}

\end{document}